\documentclass[12pt,preprint]{aastex}  

\makeatletter
\renewcommand{\@make@caption@text}[2]{%
  \begin{center}
    \makebox[\textwidth]{\rmfamily#1\quad#2}
  \end{center}
}%
\makeatother

\usepackage{xspace}
\usepackage{amssymb}
\usepackage{times}
\usepackage{verbatim}
\usepackage{color}

\newcommand{\mone}  {^{-1}}
\newcommand{\mtwo}  {^{-2}}
\newcommand{\mthree}{^{-3}}

\newcommand{\cmmtwo}{\,\mathrm{cm\mtwo}}

\newcommand{\cmmthree}{\,\mathrm{cm\mthree}}

\newcommand{\cps}   {\,\mathrm{counts\,s\mone}}

\newcommand{\eflux} {\,\mathrm{erg\,cm\mtwo\,s\mone}}

\newcommand{\kev}   {\,\mathrm{keV}}
\newcommand{\kpc}   {\,\mathrm{kpc}}

\newcommand{\kms}   {\,\mathrm{km\,s\mone}}
\newcommand{\ks}    {\,\mathrm{ks}}

\newcommand{\lum}   {\,\mathrm{erg\,s\mone}}
\newcommand{\mang}  {\,\mbox{{\AA}}\xspace}
\newcommand{\mmang}  {\,\mbox{{m\AA}}\xspace}

\newcommand{\mk}    {\,\mathrm{MK}}
\newcommand{\msun}  {\,M_\odot}

\newcommand{\pflux} {\,\mathrm{photons\,cm\mtwo\,s\mone}}

\newcommand{\arf}   {{ARF}\xspace}

\newcommand{\chan}  {{\it Chandra}\xspace}

\newcommand{\ciao}  {{CIAO}\xspace}

\newcommand{\cxo}   {{\it CXO}\xspace}
\newcommand{\heg}   {{HEG}\xspace}
\newcommand{\hetgs} {{HETGS}\xspace}
\newcommand{\hetg}  {{HETG}\xspace}
\newcommand{\letgs} {{LETGS}\xspace}
\newcommand{\meg}   {{MEG}\xspace}
\newcommand{\rgs}   {{RGS}\xspace}
\newcommand{\rmf}   {{RMF}\xspace}
\newcommand{\xmm}   {{\it XMM-Newton}\xspace}
\newcommand{\wrsix} {{WR$\,6$}\xspace}

\newcounter{ion} \newcommand{\eli}[2]{\setcounter{ion}{#2}#1{~\sc\roman{ion}}}

\shorttitle{X-ray spectra of WR~6}
\shortauthors{Huenemoerder et al.}

\begin{document}

\title{Probing Wolf-Rayet Winds: \chan/\hetg X-Ray Spectra of WR~6}

\author{David P.\ Huenemoerder\altaffilmark{a},
  K.~G.~Gayley\altaffilmark{b},
  W.-R.~Hamann\altaffilmark{c},
  R.~Ignace\altaffilmark{d},
  J.~S.~Nichols\altaffilmark{e},
  L.~Oskinova\altaffilmark{c},
  A.~M.~T.~Pollock\altaffilmark{f,g},
  N.~S.~Schulz\altaffilmark{a},
  T.~Shenar\altaffilmark{c}
\ \\
}

\altaffiltext{a} {Massachusetts Institute of Technology, Kavli
  Institute for Astrophysics and Space Research, 70 Vassar St.,
  Cambridge, MA, 02139, USA (\url{dph@space.mit.edu})}

\altaffiltext{b} {Department of Physics and Astronomy, University of
  Iowa,Iowa City, IA 52242, USA (\url{ken.gayley@gmail.com})} 

\altaffiltext{c} {Institut  f\"ur  Physik  und  Astronomie,
  Universit\"at  Potsdam,Karl-Liebknecht-Str.  24/25, D-14476 Potsdam,
  Germany (\url{wrh@astro.physik.uni-potsdam.de,
    lida@astro.physik.uni-potsdam.de, shtomer@astro.physik.uni-potsdam.de})} 

\altaffiltext{d} {Department of Physics  and  Astronomy,  East
  Tennessee  State  University, Johnson City, TN 37614, USA
  (\url{ignace@mail.etsu.edu})}  

\altaffiltext{e} {Harvard-Smithsonian Center for Astrophysics, 60
  Garden St.,MS 34, Cambridge, MA 02138, USA
  (\url{jnichols@cfa.harvard.edu})}  

\altaffiltext{f} {European Space Agency, ESAC, Apartado 78, 28691 Villanueva de la
        Ca\~{n}ada, Spain}

\altaffiltext{g}{Department of Physics and Astronomy, Hounsfield Road, University
        of Sheffield, Sheffield S3 7RH, England}

\begin{abstract}
  With a deep \chan/\hetgs exposure of \wrsix, we have resolved
  emission lines whose profiles show that the X-rays originate from a
  uniformly expanding spherical wind of high X-ray-continuum optical
  depth.  The presence of strong helium-like forbidden lines places the
  source of X-ray emission at tens to hundreds of stellar radii from
  the photosphere.  Variability was present in X-rays and simultaneous
  optical photometry, but neither were correlated with the known
  period of the system or with each other.  An enhanced abundance of
  sodium revealed nuclear processed material, a quantity related to
  the evolutionary state of the star.  The characterization of the
  extent and nature of the hot plasma in \wrsix will help to pave the
  way to a more fundamental theoretical understanding of the winds and
  evolution of massive stars.
\end{abstract}



\section{Introduction}

Within the Wolf-Rayet (WR) class are some of the most massive and
luminous of stars.  Rapid outflow of their dense stellar winds
enriches and energizes the interstellar medium before this brief phase
culminates in a core-collapse supernova detonation
\citep{Langer:2012}.  The WR stars are thus important contributors to
galactic feedback of nuclear-processed matter, mechanical energy, and
ionizing radiation throughout cosmic history, greatly affecting their
host star cluster as well as an entire galaxy.  Characterization of WR
star properties---especially mass-loss rate and composition---is
usually done through optical and UV spectroscopy
\citep{Hamann:Grafener:Liermann:2006, Hillier:Miller:1998}.  However,
structure in highly supersonic winds will invariably lead to shocks
and X-ray emission, so the X-ray regime is crucial for understanding
the nature of the wind hydrodynamics, and the structures it produces.
This also relates to the potential importance of magnetization in some
winds.  Such structures in hot-star winds can include embedded wind
shocks, magnetic confinement in some cases, or wind-wind collisions in
binary sytems \citep{Guedel:Naze:2009}.  X-ray emission lines are key
diagnostics of the high-energy processes, since line strengths and
profile shapes provide detailed information about the wind structure
and dynamics.
  There have been a large number of empirical and theoretical
  investigations of X-ray line fluxes and profiles since the advent of
  high-resolution spectroscopy with \chan and
  \xmm. \citet{Waldron:Cassinelli:2007} performed an empirical study
  of line characteristics in a collection of OB-stars.
  \citet{Herve:al:2012} examined theoretical profiles for emission
  distributed over a range of radii with different plasma conditions,
  while \citet{Leutenegger:al:2006} investigated the effects of
  distributed emission specifically for He-like lines.
  \citet{Ignace:Gayley:2002} and \citet{Leutenegger:al:2007} studied
  effects of resonant scattering on line
  profiles. \citet{Oskinova:Feldmeier:Hamann:2006} and
  \citet{Leutenegger:al:2013} looked into effects of clumping or
  porosity on line shapes. \citet{Cohen:Wollman:al:2014} did a
  systematic re-analysis of O-star line profiles to determine mass
  loss rates, \citet{Cohen:Li:al:2014} applied an underlying shock
  cooling model to determine wind structure.  \citet{Ignace:2015} has
  given a review of X-ray line profiles.  This a highly selective and
  by no means exhaustive collection of X-ray emission line modeling
  and analyis work and indicates the importance and interest 
  in this area.

The winds of hot stars are presumed to be accelerated by line-driven
radiation pressure \citep{CAK, Friend:Castor:1983,
  Pauldrach:Puls:Kudritzki:1986} which, due to instabilities, leads to
soft X-ray ($\sim0.2\kev$) emitting shocks in the acceleration zone,
typically well within a few stellar radii of the photosphere
\citep{Lucy:White:1980, Owocki:Castor:Rybicki:1988,
  Feldmeier:Puls:Pauldrach:1997, Krticka:Feldmeier:al:2009}.  The
rapid expansion leads to broad emission lines
\citep{Macfarlane:al:1991, Ignace:2001, Owocki:Cohen:2001}.  The
destruction of Helium-like forbidden line emission through intense
photospheric UV photoexcitation provides a valuable diagnostic of the
location of X-ray emission \citep{Blumenthal:Drake:al:1972,
  Waldron:Cassinelli:2001, kahn2001}.

While the above scenario is quite successful in describing the general
characteristics of OB-star X-ray emission, various details are still
under intense debate, such as the minimum radius and extent of X-ray
emission, the fraction of X-ray emitting plasma, and the clumpiness of
the wind.  There are cases, however, where this scenario is too
simple, such as in magnetically confined wind shocks
\citep{Babel:Montmerle:1997, Gagne:Oksala:al:2005}, which may have
narrower lines and plasmas dominated by high temperatures
($\sim2\kev$), or in colliding wind binaries
\citep{Luo:McCray:al:1990, Stevens:Blondin:al:1992,
  Parkin:Pittard:al:2014} which can have a range of temperatures and
profiles, depending on the orbital separation and geometrical aspect
\citep{Henley:al:2003}.

The massive winds of WR stars are also believed to host embedded
shocks \citep{Gayley:Owocki:1995} and should thus have some X-ray
characteristics in common with O-stars.  WR star winds, however, do
have some significant differences from O-stars.  Hydrogen is highly
depleted or even absent.  The mass loss rates are very high, so with
similar wind velocities to O-stars of $\sim 1000\kms$, WR stars have
much denser winds, and given their enhanced metallicity, much higher
X-ray continuum opacity.  Metallicity has a very strong effect on the
strength of WR star winds \citep{Ignace:Oskinova:1999,
  Crowther:Hadfield:2006, Grafener:Hamann:2008}.  Through the
radiative losses of emission lines, metallicity will also affect the
thermal structure of winds.

%
\begin{deluxetable}{cc}
  \tablecolumns{2}
  \tablewidth{0.45\columnwidth}
  \tablehead{
    \colhead{Property}&  \colhead{Value}
  }
  \tablecaption{\wrsix Basic Properties\label{tbl:starparams}}
  \startdata
  Spectral Type&
  WN4\\
  $R_\ast/R_\odot$&
  $2.65$\tablenotemark{a}\\
  $T_\ast\,\mathrm{[kK]}$&
  $89$\tablenotemark{a}\\
  $M [\msun]$&
  $19$\tablenotemark{a}\\
  $\log{(L_\mathrm{bol}}/{L_\odot})$&
  $5.6$\tablenotemark{a}\\
  $v_\infty[\kms]$&
  $1700$\tablenotemark{a}\\
  &
  $1950 (20)$\tablenotemark{b}\\
  $\dot{M}\,[M_\odot\,\mathrm{yr\mone}]$&
  $5\times10^{-5}$\tablenotemark{a}\\  
  %
  %
  %
  $ d\,\mathrm{[kpc]}$&
  $1.82$\tablenotemark{c}  \\
  Epoch [JD]&
  $2\,443\,199.53$\tablenotemark{d}  \\
  Period [day]&
  $3.7650$\tablenotemark{d}  \\
  \enddata
  \tablenotetext{a}{\citet{Hamann:Grafener:Liermann:2006}}
  \tablenotetext{b}{This work ($1\sigma$ uncertainty in parenthesis.)}
  \tablenotetext{c}{\citet{Howarth:Schmutz:1995}}
  \tablenotetext{d}{\citet{Georgiev:al:1999}}
\end{deluxetable}
\wrsix (EZ~CMa) has the spectral type WN~4, and is visually bright
with a magnitude $V=6.9$ \citep{vanderHucht:2001}, at a distance of
$1.8\kpc$ \citep{Howarth:Schmutz:1995}.  Its atmosphere and wind are
dominated by helium, with no detectable trace of hydrogen
\citep{Hamann:Koesterke:1998}.  A remarkable characteristic of \wrsix
is that it has a well established and consistent photometric period  of $3.7650\,$d 
  determined from $V$, $b$, $B$, and narrow band photometry, as well
  as from spectro-photometry,
though the modulation itself is highly variable in
amplitude and phase \citep{Georgiev:al:1999, Robert:al:1992,
  Lamontagne:al:1986}.  
  The photometric amplitude has ranged from about $0.1\,\mathrm{mag}$
  to unmodulated.
The star is generally considered to be single, based on its radio and
X-ray properties \citep{Dougherty:Williams:2000, Oskinova:2005,
  Skinner:Zhekov:al:2002b}.
  \citet{Georgiev:al:1999} argued for binarity based upon their
  determination that variations are coherent in phase over long times.
  \citet{Morel:St-Louis:Marchenko:1997}, on the other hand, argued for
  a single star, with variability due to structure in the wind, since
  the variability sometimes vanishes.  \citet{Skinner:Zhekov:al:2002b}
  rejected the hypothesis of a compact companion, but they could not
  exclude the possibility of a low-mass pre-main-sequence companion,
  though they concluded that without direct evidence, the star should
  continue to be considered single.  Hence, for the present purposes,
  we will consider \wrsix to be single.

\wrsix has no detected global magnetic field
\citep{delaChevrotiere:2013}, and so is not suspected of having a
magnetically confined wind.

Some fundamental parameters of \wrsix are given in
Table~\ref{tbl:starparams}, many from the comprehensive modeling of
galactic WN stars by \citet{Hamann:Grafener:Liermann:2006}, or as
updated in this work.

The first high-resolution X-ray spectrum of \wrsix, obtained with the
\xmm Reflection Grating Spectrometer (\rgs), showed that the X-rays
from this star are generated by a new and unknown mechanism, different
from O-star winds \citep{Oskinova:al:2012} in that the X-rays do not
appear to be consistent with O-star embedded wind shocks
\citep{Krticka:Feldmeier:al:2009}, or with magnetic confinement close
to the star.  The hot, X-ray emitting plasma of \wrsix exists far out
in the wind, as determined from He-like line ratios and also from the
presence of Fe~K fluorescence, likely produced from cool wind plasma
illuminated by hard X-rays.  New mechanisms are required to explain
this emission.  This fascinating situation demonstrates that we do not
fully understand some important aspects of the winds of massive stars
in the crucial evolutionary stages immediately prior to a supernova
explosion.

The wind-broadened line profiles in \wrsix were only marginally
resolved by the \xmm\ \rgs.  Therefore we obtained Chandra spectra
using the High Energy Transmission Grating Spectrometer (\hetgs) which
provides significantly superior spectral resolving power, being four
times better than that of the \rgs. The \hetgs pass-band covers the
strategic lines of S, Si, Mg, and Ne, which are needed to understand
the distribution of the hottest gas in the WR wind---He-like ratios
reveal the radial location, H-to-He-like ratios specify temperatures,
and line positions and profile widths reveal the wind dynamics.
Therefore, we used \hetgs to study this mystery uncovered by \xmm.
Figure~\ref{fig:summaryspec} gives an overview of the important part
of the spectrum observable with \hetgs.

\section{Observations and Calibration}

\begin{figure}[!htb]
  \centering\leavevmode
  \includegraphics[width=0.80\columnwidth, viewport=0 0 414 225]{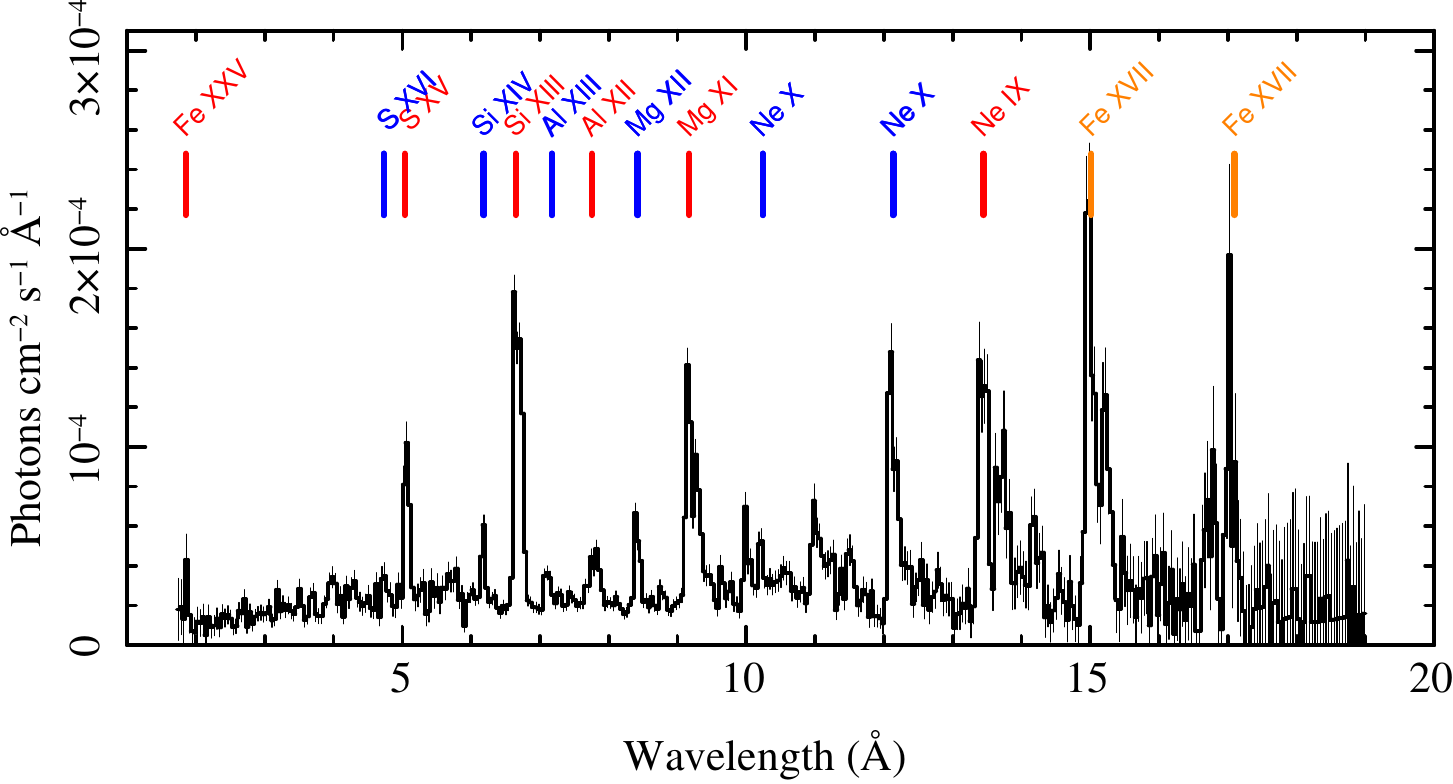}\\
  \includegraphics[width=0.80\columnwidth, viewport=0 0 414 230]{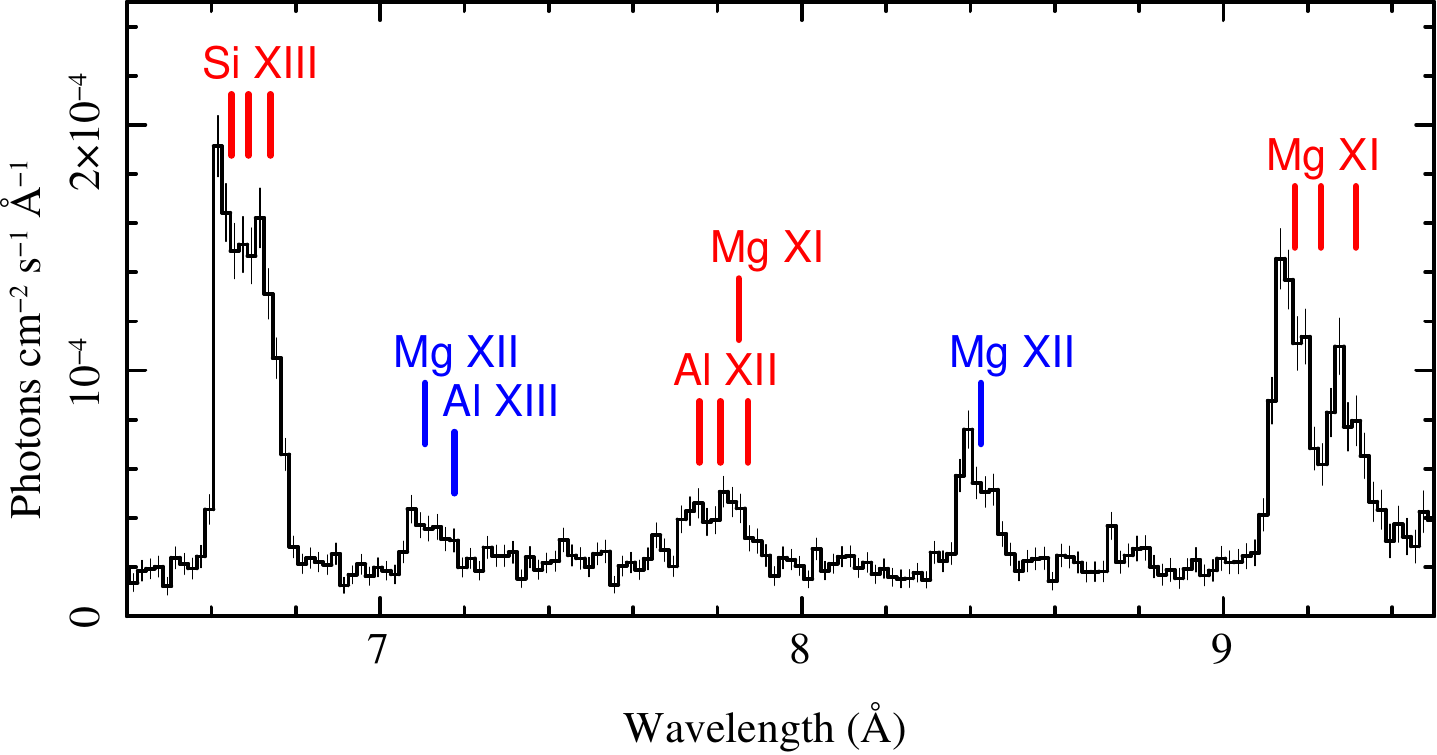}\\
  \caption{The \hetgs spectrum of \wrsix: combined \heg and \meg first
    orders, in flux units.  The locations of some spectral features,
    primarily H- and He-like ions, are marked. The top panel shows the
    useful range of the instrument, with bins of $0.04\mang$ (2 \meg,
    8 \heg resolutions elements).  The bottom panel shows detail,
    including the resonance, intercombination, and forbidden line
    locations (short-to-long wavelength) for a narrow region, binned
    to $0.04\mang$. Line labels are blue for H-like ions, red for
    He-like, and orange for others.  Error bars are $1\sigma$.}
  \label{fig:summaryspec}
\end{figure}
Using the \chan/\hetg spectrometer \citep{HETG:2005}, we observed
\wrsix three times in 2013 for a total of $440\ks$. Dataset
identifiers and exposure times are given in Table~\ref{tbl:obsparams}.
The \hetgs spectra cover the range from about $1$--$30\mang$, as
dispersed by two types of grating facets, the High Energy Grating
(\heg) and the Medium Energy Grating (\meg), with resolving powers
ranging from 100 to 1000, with approximately constant
full-width-half-maxima ({\em FWHM}) of $12\mmang$ for \heg and
$23\mmang$ for \meg.

The \chan data were reprocessed with standard Chandra Interactive
Analysis of Observations (\ciao) programs \citep{CIAO:2006} to apply
the most recent calibration data (\ciao 4.6 and the corresponding
calibration database, version 4.5.9).  The counts spectra are thus
composed of 4 orders per source per observation: the positive and
negative first orders for each grating type, the \meg and \heg, which
have different efficiencies and resolving powers.  The default binning
over-samples the instrumental resolution by about a factor of 4.

Several calibration files are required for analysis to convolve a
model flux spectrum with the instrumental response in order to produce
model counts.  These are made for each observation and each spectral
order by the \ciao programs which use observation-specific data in
conjunction with the calibration files to make the effective area
files (``Auxiliary Response File'', or \arf) and the spectral
redistribution and extraction-aperture efficiency files (``Response
Matrix File'', or \rmf) \citep{DavisJE:2001b}.

Since emission lines in the \wrsix spectrum are broad and well sampled
at the lower \meg resolution, during analysis we regridded the \heg
spectra and response matrices onto the \meg grid so that the spectra
could be combined.  This primarily aids visualization of the data as
one spectrum.

Figure~\ref{fig:summaryspec} shows the combined photon flux
spectrum\footnote{Flux calibration is described in
  Appendix~\ref{app:fluxcal}.} for the \heg and \meg first orders.
Some prominent features are marked, mainly the H-like and He-like
lines of abundant chemical elements, and some strong Fe lines.  Due to
the low effective area beyond about $20\mang$, the interstellar medium
(ISM) absorption on the line-of-sight, wind absorption (intrinsic to
\wrsix), {and the absence of oxygen in the atmosphere}, there
is little signal {detected by the \hetgs} at the longer
wavelengths.  The ISM component has a transmission factor of $0.95$ at
$5\mang$, but $\sim0.2$ at $20\mang$.  The wind-local absorption has a
slightly larger effect, with a combined ISM/wind result of about 50\%
transmission at $10\mang$, but only 3\% at $20\mang$.  (The wind-local
model of emission and absorption is discussed in more detail in
Section~\ref{sec:specmodel} and Appendix~\ref{app:specmodel}.)

\begin{figure}[!htb]
  \centering\leavevmode
  \includegraphics[width=0.45\columnwidth]{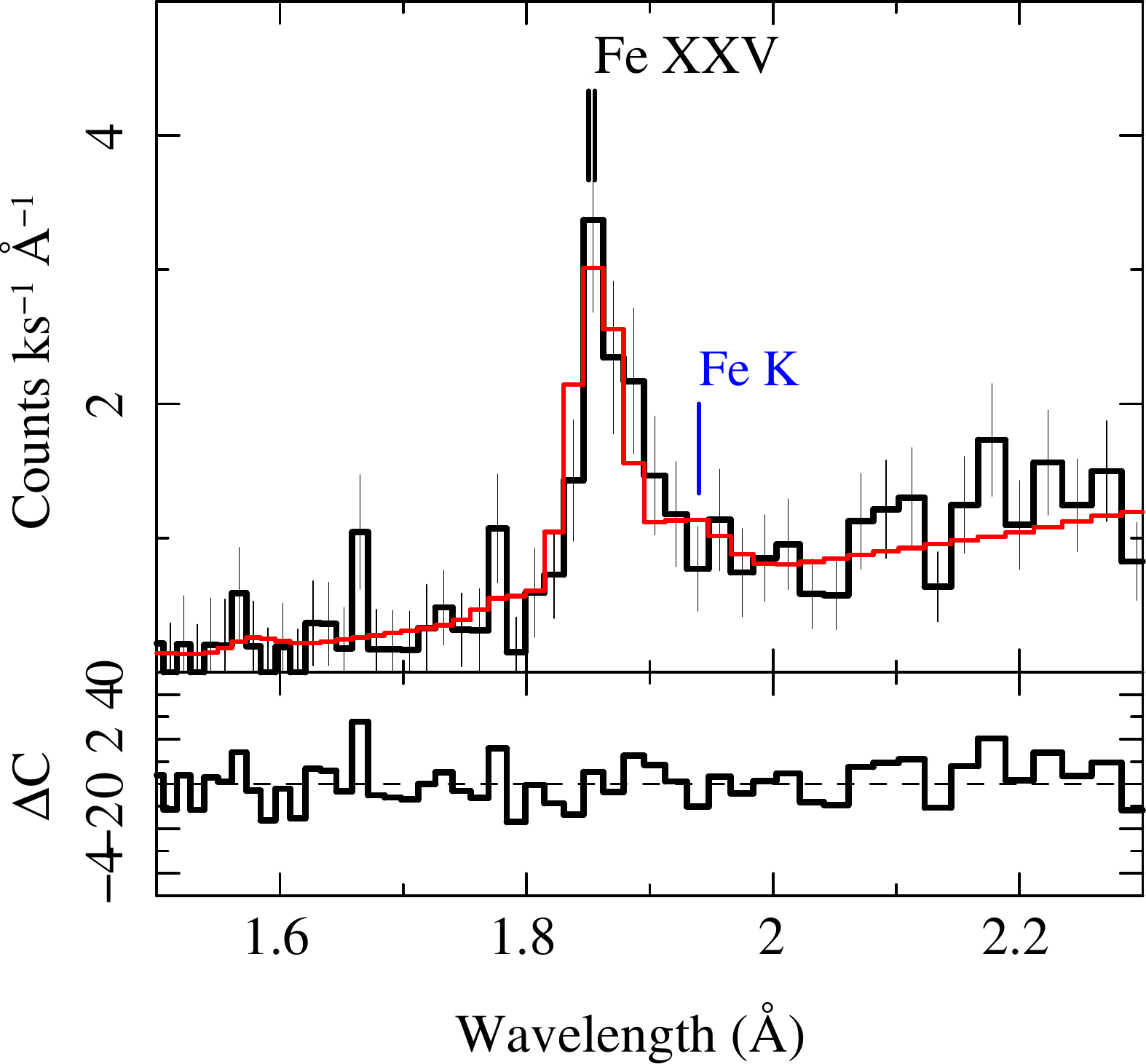}\quad
  \includegraphics[width=0.45\columnwidth]{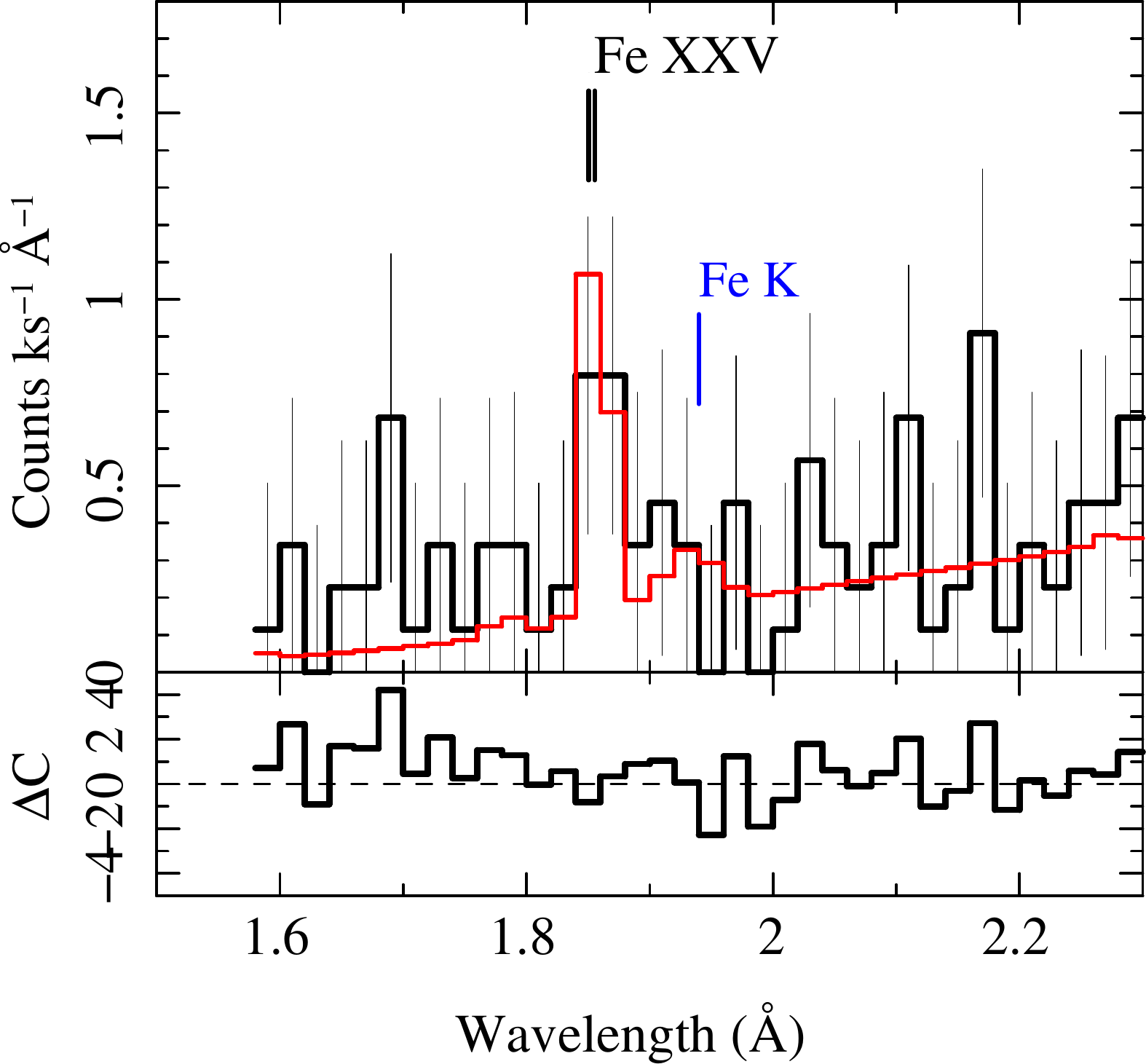}
  \caption{The \hetgs zeroth order spectrum of \wrsix (left) and the
    dispersed spectrum (right) in the Fe~K region. The strong emission
    line is the blend of He-like resonance, intercombination, and
    forbidden lines of \eli{Fe}{25} ($1.85\mang, 6.70\kev$) having
    peak emissivity near $60\mk$. The model shown in red is an
    isothermal APEC model with $T=43\mk$.  A fit of an Fe~K
    fluorescent feature at $1.94\mang$ ($6.39\kev$) is consistent with
    the \xmm-determined value \citep{Oskinova:al:2012}, though is of
    very low significance in these data, changing the statistic from
    $1.12$ to $1.04$.  Error bars are $1\sigma$.  }
   \label{fig:zofek}
\end{figure}

The zeroth order is useful for measuring the \eli{Fe}{25} line,
constraining the hottest plasmas present, and for variability studies,
since the CCD photon pileup being fairly small at a rate per frame of
less than $0.1\cps$. We show this region of the zeroth order spectrum
in Figure~\ref{fig:zofek}.

%
\begin{deluxetable}{crrc}
  \tablecolumns{4}
  \tablewidth{0.0\columnwidth}
  \tablecaption{Observational Information\label{tbl:obsparams}}
  \tablehead{
    \colhead{Date\tablenotemark{a}}&
    \colhead{ObsID\tablenotemark{b}}&
    \colhead{Exp}&
    \colhead{$\langle Rate\rangle$\tablenotemark{c}}\\
    &
    &
    \colhead{$[\mathrm{ks}]$}&
    \colhead{$\mathrm{[cts\ks\mone]}$}
  }
  \startdata
  2013-05-03T04:34:39& 14534 & 168& 51.2 (0.6)\\
  2013-08-15T08:22:51& 14535 & 97&  55.7 (0.8)\\
  2013-08-19T13:45:08& 14533 & 175& 57.9 (0.6)\\ \tableline\tableline
  Total exposure [ks]&&  440\\
  Count rate (diffracted)\tablenotemark{c} $\mathrm{[cts\, ks\mone]}$& $32.8$\\
  Count rate (direct)\tablenotemark{c} $\mathrm{[cts\, ks\mone]}$& $22.4$\\
  \enddata
  \tablenotetext{a}{The date of observation is given by the FITS
    standard {\tt CCYY-MM-DDThh:mm:ss.sss}, which is the universal time
    at the start of the exposure,  as encoded into the FITS file header
    keyword {\tt  DATE-OBS}.}
  \tablenotetext{b}{The \chan observation identifier.}
  \tablenotetext{c}{The mean count rate is given for each ObsID for the
    combined first and zeroth orders, with the one standard deviation
    error of the mean in parentheses.  The diffracted count rate is
    for photons in the first orders of \heg and \meg.  The direct rate
    is for photons in the zeroth order.}
\end{deluxetable}

\section{Spectral Modeling\label{sec:specmodel}}

The X-ray emission line strengths and profiles from stellar winds are
very sensitive to the wind structure and dynamics.
\citet{Macfarlane:al:1991} showed that the profile from a
geometrically thin shell ranges from a flat-topped profile of width
determined by the expansion velocity, to a triangular profile if the
foreground wind shell's continuum opacity obscures the receding shell.
\citet{Ignace:2001} studied the line profile for a shell undergoing
constant spherical expansion, including the limit of high continuum
opacity.  \citet{Owocki:Cohen:2001} extended the formalism to a
radially-dependent expansion velocity with arbitrary continuum opacity
and showed that profiles are generally asymmetric, with a blue-shifted
centroid, again due to absorption of the receding wind by the
intervening wind.  They also modeled the effect of a minimum radius of
line formation, showing how a central void (in both velocity and
emission measure) can strongly affect the line profile by flattening
the peak, since the volume of higher density, low-velocity plasma has
been reduced.  \citet{Oskinova:Feldmeier:Hamann:2004,
  Oskinova:Feldmeier:Hamann:2006} showed how clumping in the wind can
affect the emergent profile, and, depending on clumping parameters,
result in unshifted, symmetric profiles, in contrast to a smooth
wind's skewed profiles.

The \xmm/\rgs spectra showed that the X-ray line profiles of \wrsix
are broadened by an amount consistent with $v_{\infty}=1700\kms$, and
that a blueshifted Gaussian profile with $\Delta v \approx -650\kms$
generally gave an adequate fit \citep{Oskinova:al:2012}.  With the
higher resolution of \hetg, we find that a Gaussian is a very poor
approximation to the profile.  Instead, we find that a spherical
constant velocity expansion model, similar to that described by
\citet{Ignace:2001}, gives a very good fit.

As a baseline for modeling the lines, we assumed a provisional plasma
model based on the \rgs analyis of \citet{Oskinova:al:2012}.  This is
primarily a phenomenological model because it does not entail
distributed emission and absorption throughout the wind, but is a slab
model, with an underlying multi-thermal plasma, overlying wind
absorption with ionized wind edges, and an interstellar foreground
absorption component.  This serves as a basis for identifying and
characterizing spectral features, and incorporates blends from the
basic plasma model.  We extended the 3-temperature APEC
\citep{Smith:01,Foster:Smith:Brickhouse:2012} model used in the \rgs
analysis by including an additional temperature component to satisfy
the higher energy response of \hetg, and we re-fit the temperatures
and normalizations simultaneously for \hetg's first and zeroth orders,
\xmm\ \rgs-1 and -2 first and second orders, and the \xmm EPIC MOS and
PN spectra.  Absorption components, which were determined from \xmm
analysis, were left fixed, since they are better constrained by the
longer wavelength data.  A model summary is given in
Table~\ref{tbl:modelparams}, and more details are given in
Appendix~\ref{app:specmodel}.
%
%
\begin{deluxetable}{cc}
  \tablecolumns{2}
  \tablewidth{0.0\columnwidth}
  \tablecaption{Model Parameters\label{tbl:modelparams}}
  \tablehead{
    \colhead{Property}&  \colhead{Value}
  }
  \startdata
  Temperatures, $T_\mathrm{x}\,\mathrm{[MK]}$&
  $1.5$, $4.0$, $8.0$, $50$\\
  Normalizations [relative]&
  $10$, $3.3$, $1.9$, $0.9$\\
  $\log f_\mathrm{x}[\eflux]$ &
  $-11.9$\\
  $\log L_\mathrm{x}[\lum]$ &
  $32.9$\\
  $\log (L_\mathrm{x}/L_\mathrm{bol})$ &
  $-6.3$\\
  $\log N_\mathrm{H}\,\mathrm{[cm\mtwo]}$\tablenotemark{a}&
  $21.2$ \\
  \enddata
  \tablenotetext{a}{Foreground interstellar absorption.  Flux is
    as-observed, whereas luminosity has been corrected for foreground
    interstellar absorption.}
\end{deluxetable}

\subsection{Line Profile Modeling}

The lines are well resolved by \hetg and are very non-Gaussian.  They
are fin-shaped, with a sharp blue edge, upward-convex, sloping down to
the red.  Figure~\ref{fig:summaryspec} (bottom panel) shows detail for
a narrow spectral region and demonstrates the near-vertical blue edge
and maximum being blueward of line center.  The profiles are very much
like those shown in \citet{Ignace:2001}, or of the profiles from
optically thick (in the continuum) winds with a large central cavity
\citep[e.g.][see their Figure 2]{Owocki:Cohen:2001}.

For WR~winds it is natural to expect line profiles that sample the
asymptotic flow, because the winds are quite dense such that optical
depth unity in photoabsorption of X-ray emission is expected to be at
relatively large radius.  The analytic solutions of
\citet{Ignace:2001} were developed for just this case.

We have adopted the analytic form of \citet{Ignace:2001} for the line
profile, $f(w_{\rm z})$, valid in the limit of large optical depth:
\begin{equation}
  f(w_{\rm z})  =  f_0  \left[ \frac{\sqrt{ 1 - w_{\rm z}^2 }} {\arccos( -w_{\rm z} )} \right]^{1+q}
  \label{eqn:lsf}
\end{equation}
where $f_0$ is a normalization constant, and
$w_{\rm z}$ is the dimensionless scaled velocity along the
line-of-sight $z$,
%
$    w_{\rm z} = ({c}/{v_{\infty}})( \lambda/\lambda_0 - 1 )   $
for a line having rest wavelength $\lambda_0$.  In this formula the
emissivity per volume is assumed to vary as the square of density.
However, an additional modification to the emissivity is allowed in
the form of $r^{-q}$, with $q > -1$
{
as outlined in \citet[][see Equation 8 and associated discussion therein]{Ignace:2001}.
}
This $q$ parameter serves to
modify the shape of the line profile from a pure density squared
result.  Physically, one can think of this accompaniment to the line
emissivity as representing perhaps a number of factors, including a
volume filling factor to accomodate clumping, or a radius-dependent
X-ray temperature distribution to accomodate variations in shock
strength.  In this sense different $q$ values are to be expected from
fits to different lines, in contrast to seeking a single value of $q$
that applies to all lines.

\begin{figure}[!htb]
  \centering\leavevmode
  \includegraphics*[width=0.33\columnwidth]{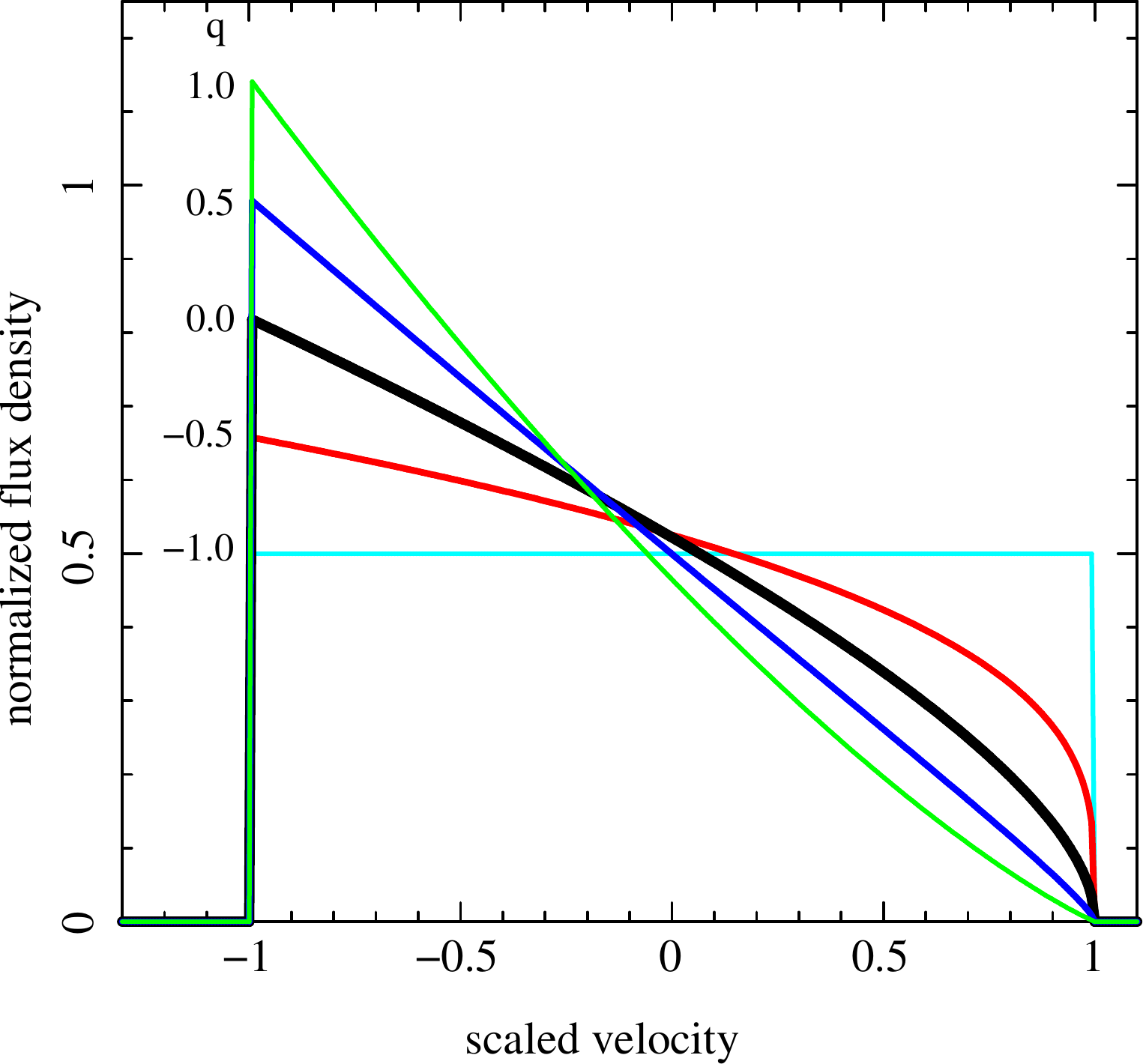}
  \caption{Example intrinsic line profiles for constant spherical expansion for
    several values of $q$, as defined by Equation~\ref{eqn:lsf}.
  }
  \label{fig:lsffun}
\end{figure}

As an illustration, a family of model line profiles are shown in
Figure~\ref{fig:lsffun} for a range of $q$ values.  Note that
the photoabsorption optical depth is assumed large, meaning that 
X-rays from near the stellar photosphere are strongly absorbed,
and that the X-ray line profile is formed generally in the vicinity of
optical depth unity in photoabsorption, although this depends in
detail on the value of $q$.  As seen in Figure~\ref{fig:lsffun},
the case of $q=0$ gives the canonical ``fin''-shaped line profile.
Positive values of $q$ serve to exacerbate the relative sharpness
of the fin; negative values of $q$ reduce, to the extreme
that $q=-1$ recovers a ``flat-top'' line profile that is normally
associated with the case of zero photoabsorption.

In using the adopted form of Equation~\ref{eqn:lsf} to model line
profiles, we have made several assumptions.  First, that the X-ray
line emission is taken to be well-described by collisional ionization
equilibrium in which every collisional excitation is from the ground
state and results in a radiative transition, with negligible optical
depth in the lines.  This serves as the basis of the density-squared
emissivity.  Second, the continuum opacity of the \wrsix wind is very
large.  In soft X-rays, the large radial optical depth prevents us
from seeing down to the acceleration zone, so constant expansion is a
reasonable assumption for the visible plasma. In this limit, if we
assume that all X-ray emission lines have the same profile, then it
means they all sample the same terminal velocity with the same
temperature (or same temperature distribution).

To relax the latter assumption, that all lines have a common thermal
origin with similar hot-gas filling factor, we allow the exponent,
$q$, to be non-zero.  In this way, we can fit individual profiles to
explore trends in expansion velocity or shape.  For example, the
continuum opacity is lower at shorter wavelengths; if it is
significantly smaller such that we can see deeper, where conditions
may be different, we might expect the shortest wavelength lines to
have a different shape from the longer wavelength lines even though
all the lines form in the asymptotic terminal speed flow.

We have implemented the model line profile as a parametric fit
function, but also as a global intrinsic line profile in the APEC
model evaluation {(that is, our APEC model, in addition to the usual
parameters of temperatures, normalizations, abundances, and Doppler
shift, has wind-profile parameters)}.  To determine the profile
parameters (since the global plasma model is not necessarily the best
model for all features), we independently fit narrow spectral regions
containing strong or important lines by adopting the 4-temperature
model as a starting point and then fit the normalization, relevant
elemental abundances (to allow optimization of the line-to-continuum
ratios), and---of primary interest---wind parameters $q$ and
$v_\infty$.  We adopted line-of-sight Doppler velocity of $46.2\kms$,
which is the exposure-time weighted mean of the systemic
\citep{Firmani:al:1980} plus line-of-sight velocities for the three
observations, which ranged from $34$--$66\kms$. The differences
between the observations are negligible in consideration of the
resolution and the line width, though very important to set a priori
because the terminal velocity and Doppler shift are degenerate
parameters.
{
  We take the Doppler velocity as a given, and do not fit the line
  center.
}

\begin{figure}[!htb]
  \centering\leavevmode
  \includegraphics*[width=0.45\columnwidth]{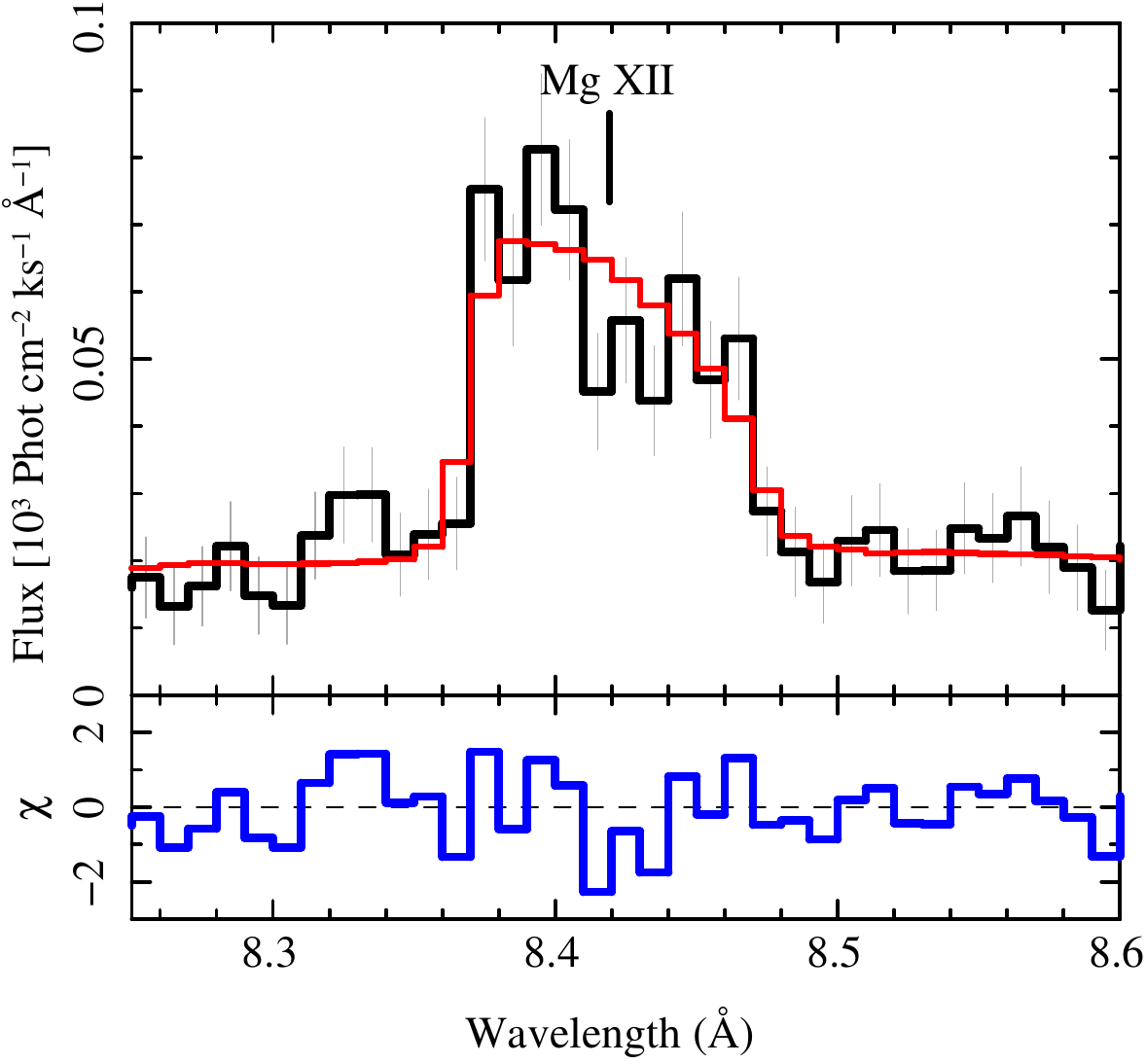}
  \includegraphics*[width=0.45\columnwidth]{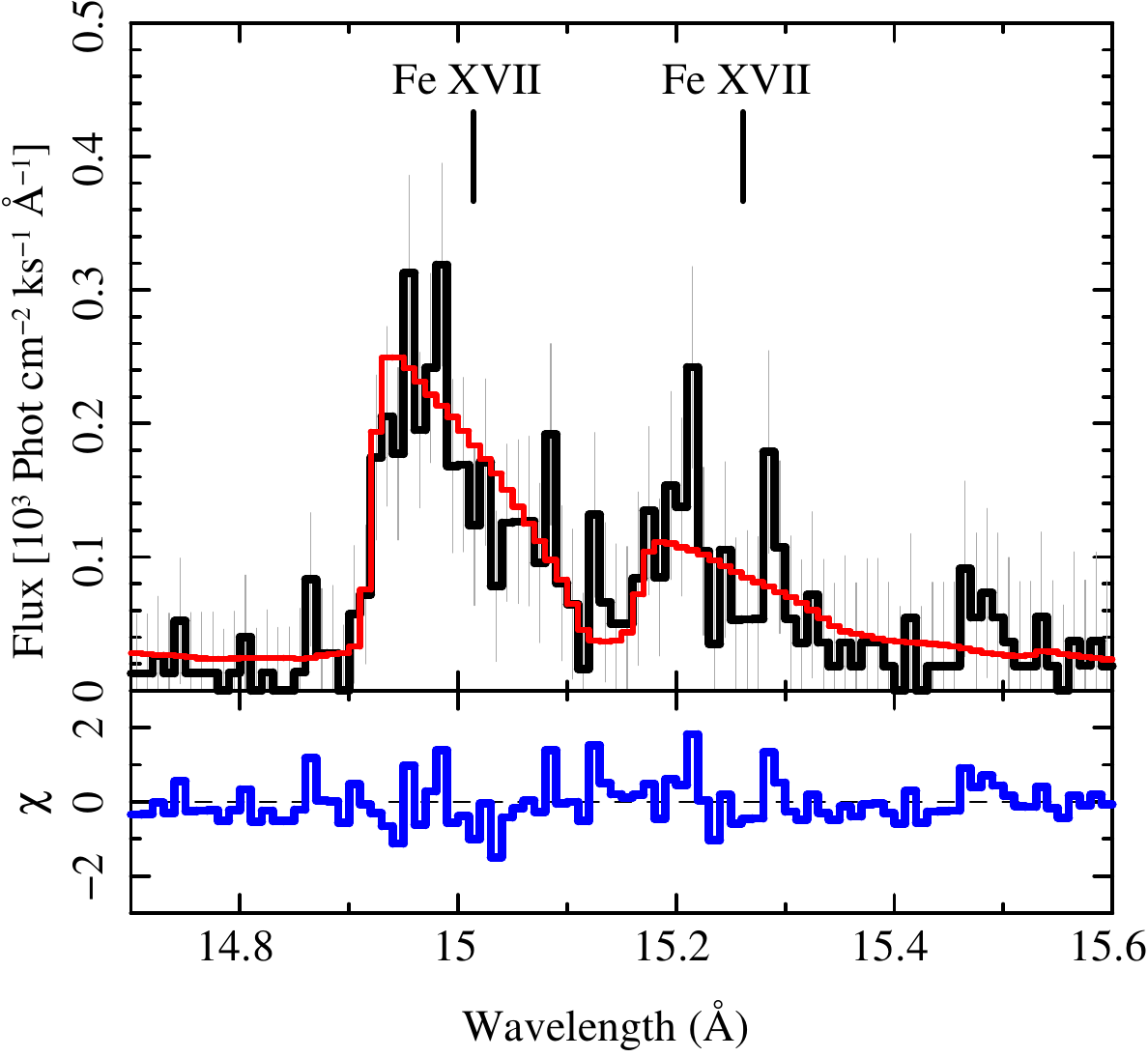}
  \caption{The fits to \eli{Mg}{12} (left) and \eli{Fe}{17} (right),
    which shows the data (black) and model profile (red) for some
    relatively unblended features to demonstrate the ``fin'' shape of
    the lines.  The line centers of the strongest lines in the region
    are marked; other model lines in the regions shown are an order of
    magnitude fainter than the brightest.  The lower panels show the
    fit residuals.  }
  \label{fig:mgprof}
\end{figure}

{
  One must be careful to distinguish the line {\em center} from a line
  {\em centroid}.  A common diagnostic of stellar winds is often
  referred to as a ``blueshifted profile''.  This is incorrect when
  referring to the volume-integrated profile: the {\em centroid} is
  blueward of line center because the wind opacity causes the emergent
  profile (centered at zero velocity) to be skewed through absorption
  of locally redshifted wind emission.  Here we specifically refer to the
  theoretical line profiles' centers when we indicate the line
  position.
}
{
  In Figure~\ref{fig:mgprof}, we show observed and model profiles for
  relatively clean portions of the spectrum, to demonstrate the
  ``fin''-shape and relative position of the line center.  The models
  were APEC plasmas fit within the narrow regions as described above.
}

For the He-like lines, we also used density as a free parameter to
allow the important forbidden ($f$) to intercombination ($i$) line
flux ratio ($R=f/i$) to be free; use of density is simply a convenient
proxy for consideration of UV photoexcitation from the forbidden to
intercombination levels \citep{Blumenthal:Drake:al:1972}.  The density
dependence was computed with APEC, then parameterized for use as a
line emissivity modifier.\footnote{See
  \url{http://space.mit.edu/cxc/analysis/he\_modifier} for details,
  data, and code.}  As an alternative, we also fit the He-like lines
parametrically using a linear combination of a continuum and line
components. This is less physically constrained than the plasma-model
approach since, for instance, there is no a priori relation between
the forbidden and intercombination line.  Uncertainties in this
approach tend to be large because the lower limit on the
intercombination line can be very small, and the
forbidden-to-intercombination ratio arbitrarily large; hence we favor
the APEC-based results.  The fitted values are given in
Table~\ref{tbl:grparametric}.

\begin{deluxetable}{rrrc}
  \tablecolumns{4}
  \tablewidth{0.0\columnwidth}
  \tablecaption{$R$ Values for He-like Lines\label{tbl:grparametric}}
  \tablehead{
    \colhead{Ion}&
    \colhead{$\lambda$}&
    \colhead{$R$}&
    \colhead{$R_0$}\\
    &
    \colhead{$\mang$}&
    \colhead{$f/i$}&
    \colhead{$\mathrm{max}(f/i)$}
  }
  \startdata
  \eli{S}{15}&  $5.04$&    $1.85$ $(0.50,    1.85)$&$1.85$\\
  \eli{Si}{13}& $6.65$&    $2.41$ $(2.19,    2.41)$&$2.41$\\
  \eli{Mg}{11}& $9.17$&    $2.95$ $(2.32,    2.95)$&$2.96$\\
  \eli{Ne}{9}& $13.45$&    $3.35$ $(2.29,    3.35)$&$3.35$
  \enddata
  \tablecomments{\ Limits for 90\% confidence intervals are given in
    parentheses.  Wavelengths are given for the resonance lines.
    $R_0$ is for the 4-temperature model, and so may differ slightly
    from the value at the temperature of maximum emissivity for each
    ion.  }
\end{deluxetable}

\begin{figure}[!htb]
  \centering\leavevmode
  \begin{tabular}{c}
  \includegraphics*[width=0.5\columnwidth,viewport=0 42 430 260]{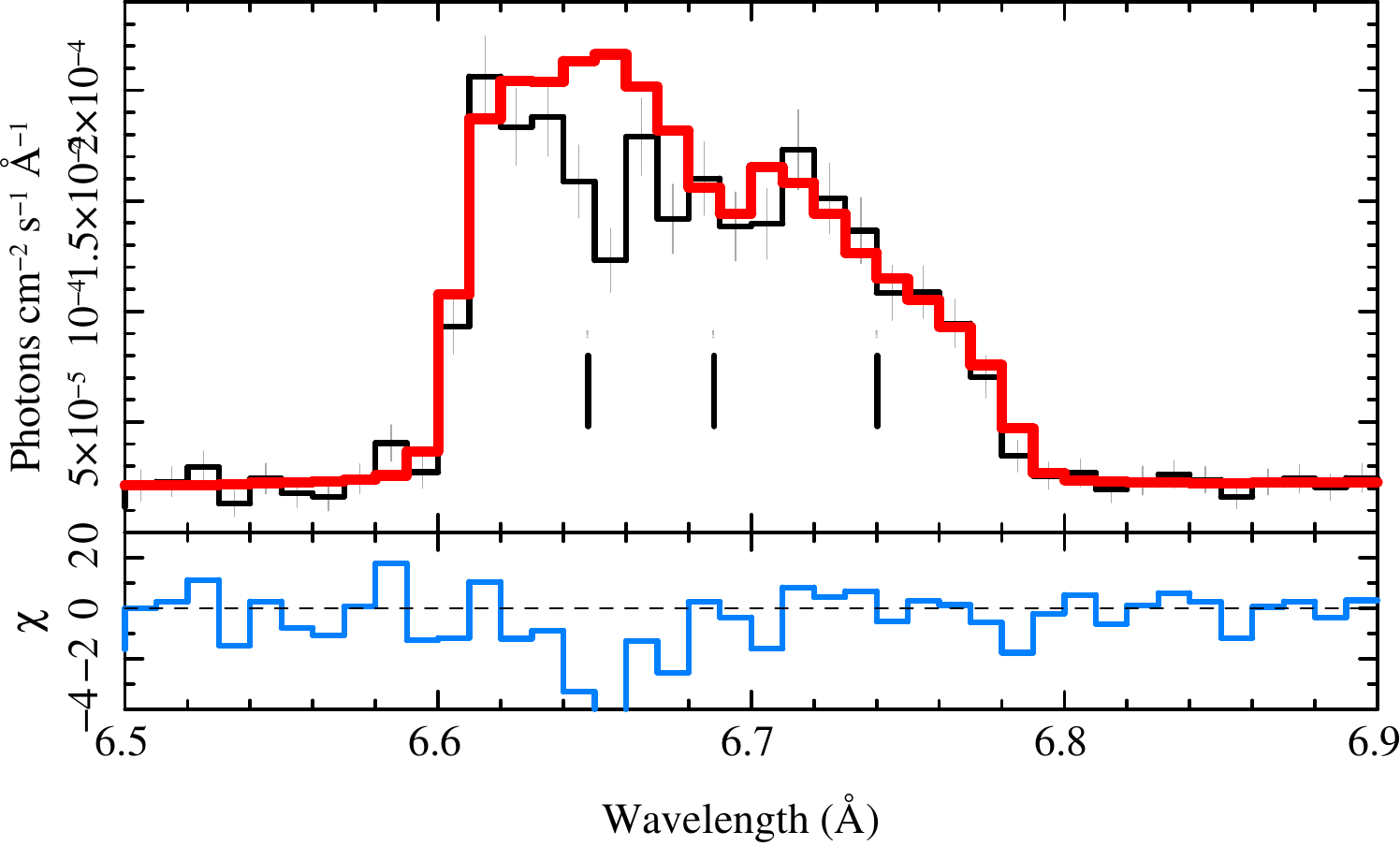}\\
  \includegraphics*[width=0.5\columnwidth,viewport=0  0 430 264]{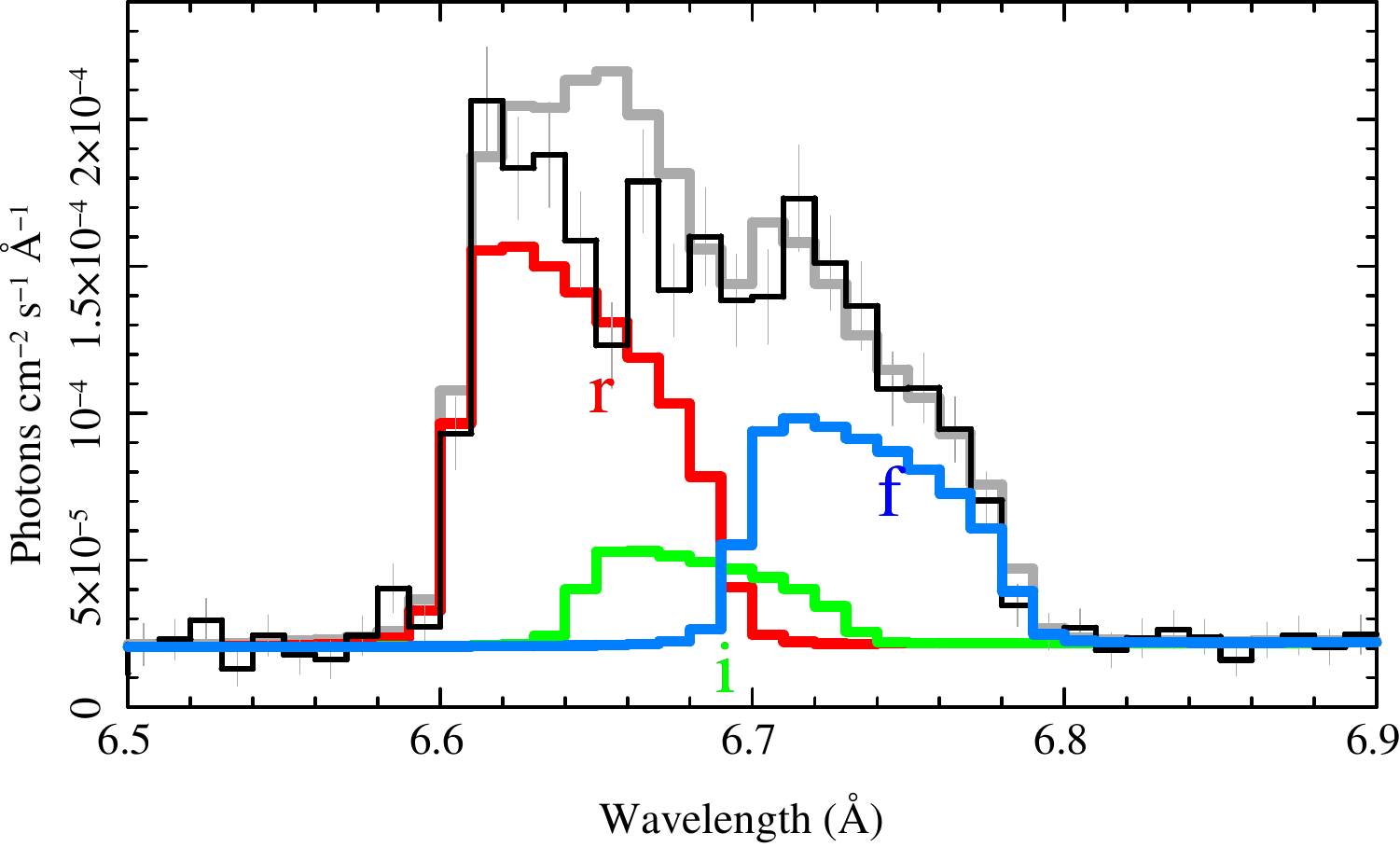}
  \end{tabular}
  \caption{The fit to \eli{Si}{13}.  The top panel shows the photon
    flux spectrum (black), the model (red) and the residuals below in
    the small sub-panel.  The bottom plot shows the same counts
    (black) and model (gray), but also shows the components for the
    resonance (red), intercombination (green), and forbidden (blue)
    lines (also labeled as $r$, $i$, and $f$).  Line {centers
      ({\em not centroids}) are marked in the upper panel.  The three
      model components shown do not sum to the total, which includes
      additional flux from many dielectronic recombination lines, as
      well as a small contribution from \eli{Mg}{12}.}  }
  \label{fig:siprof}
\end{figure}
We show as an example the fit to the \eli{Si}{13} lines in
Figure~\ref{fig:siprof}, as well as the decomposition into the
component profiles.  The lines are well resolved.  One discrepancy for
\eli{Si}{13} is a relatively large feature in the residuals, which
might mean that the $i$-component is narrower than the others (all
three components shared the same profile parameters in the fit, as
shown in Figure~\ref{fig:qvvsw} and Table~\ref{tbl:qvinf}).

\begin{figure}[!htb]
  \centering\leavevmode
  \includegraphics[width=0.8\columnwidth]{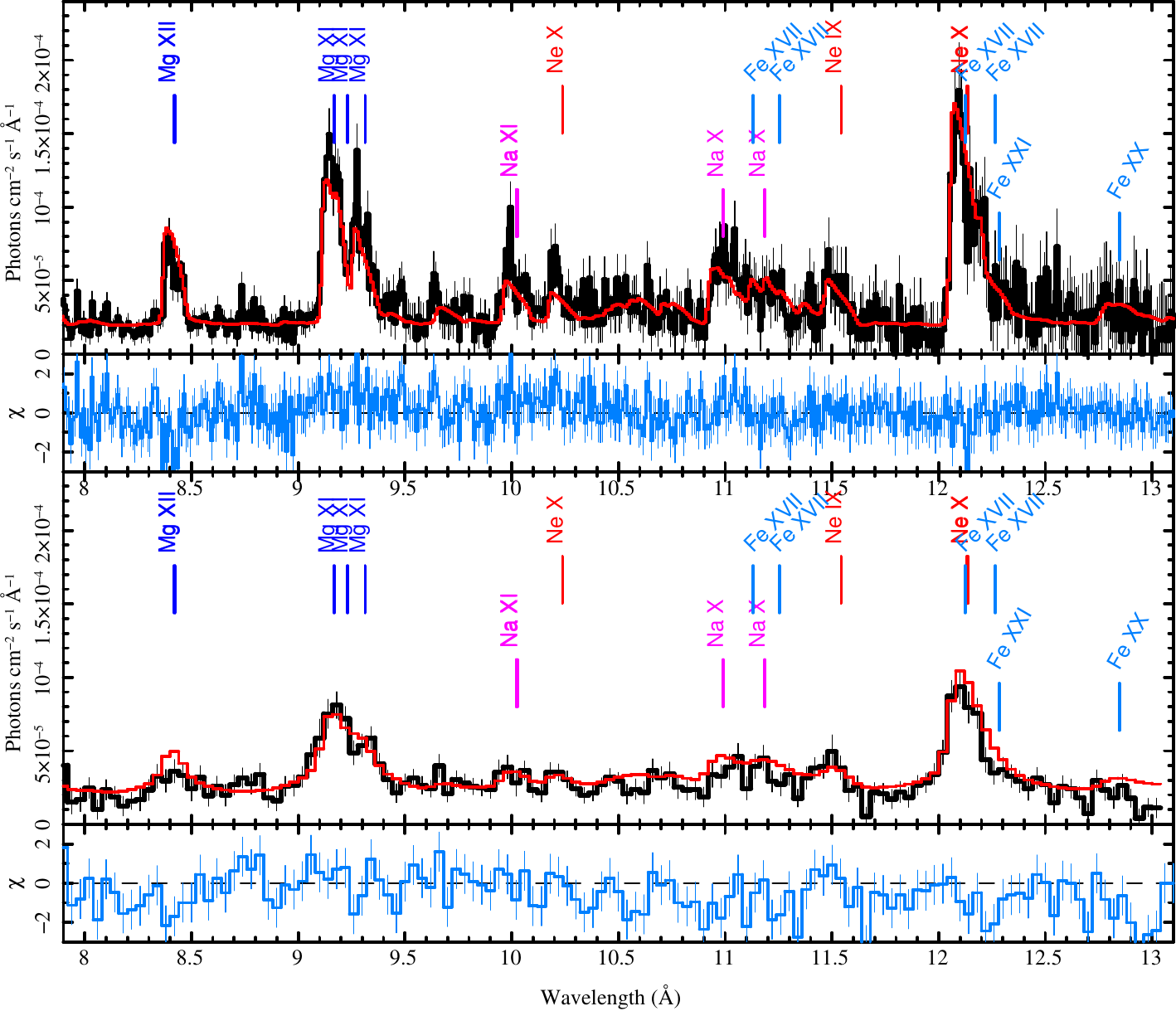}
  \caption{Here we compare the model to the photon flux \hetg
    spectrum (top), and the same model to the \rgs spectrum.  In each
    case, the black histogram is the observed spectrum
    with two bins per resolution element, and the red is the  model.
    Below each are the residuals.  In this region it is clear that the
    \hetg has resolved the sharp blue edges on the profiles, which
    were not evident in the lower-resolution \rgs data.
  }
  \label{fig:hetgvsrgs}
\end{figure}

In Figure~\ref{fig:hetgvsrgs} we show a broader region of the spectrum
along with the same spectral region as observed with \rgs (bottom
panel pair).  This clearly demonstrates the character of the profiles'
vertical blue edge and the great advantage of the \hetg's higher
resolution for determining the profile shape.  With \rgs, we could
determine that the lines are broad, but a near-Gaussian profile was
sufficient to fit them.

From fitting the narrow spectral regions {with the APEC-based model
including the model line profile of Equation~\ref{eqn:lsf},} we have determined an
error-weighted-mean expansion velocity of $1950\pm20\kms$, somewhat
larger than the 
%
{
  average value of $1700\kms$ as determined by
  \citep{Hamann:Grafener:Liermann:2006} from analysis of the full
  spectrum, but smaller than the maximum value of $2100$--$2500\kms$
  as observed among the UV lines in the several hundred archived IUE
  spectra.
}
Our result is dominated by the best-determined value for \eli{Si}{13}.
A straight mean and standard deviation gives $1880\pm140\kms$.  The
lines are all consistent with a shape parameter of $q\simeq-0.2$
{
  \citep{Ignace:2001}, which is close to zero, the nominal value for
  density-squared emissivity and uniform expansion (see
  Equation~\ref{eqn:lsf} and Figure~\ref{fig:lsffun}).  }
We show the determinations for each feature measured in
Figure~\ref{fig:qvvsw}, and the values are listed in
Table~\ref{tbl:qvinf}.

\begin{figure}[!htb]
  \centering\leavevmode
  \includegraphics[width=0.40\columnwidth]{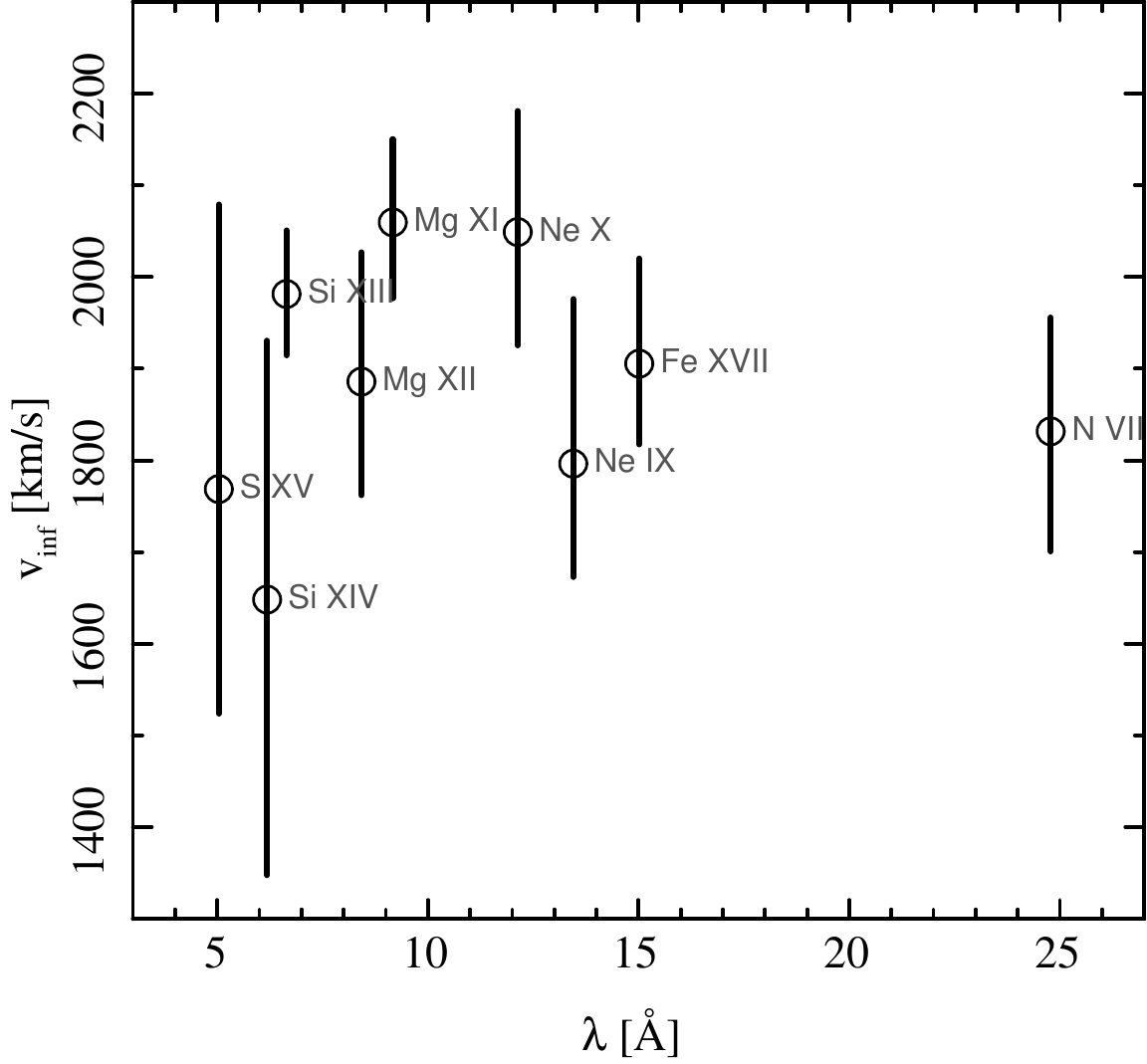}\qquad
  \includegraphics[width=0.40\columnwidth]{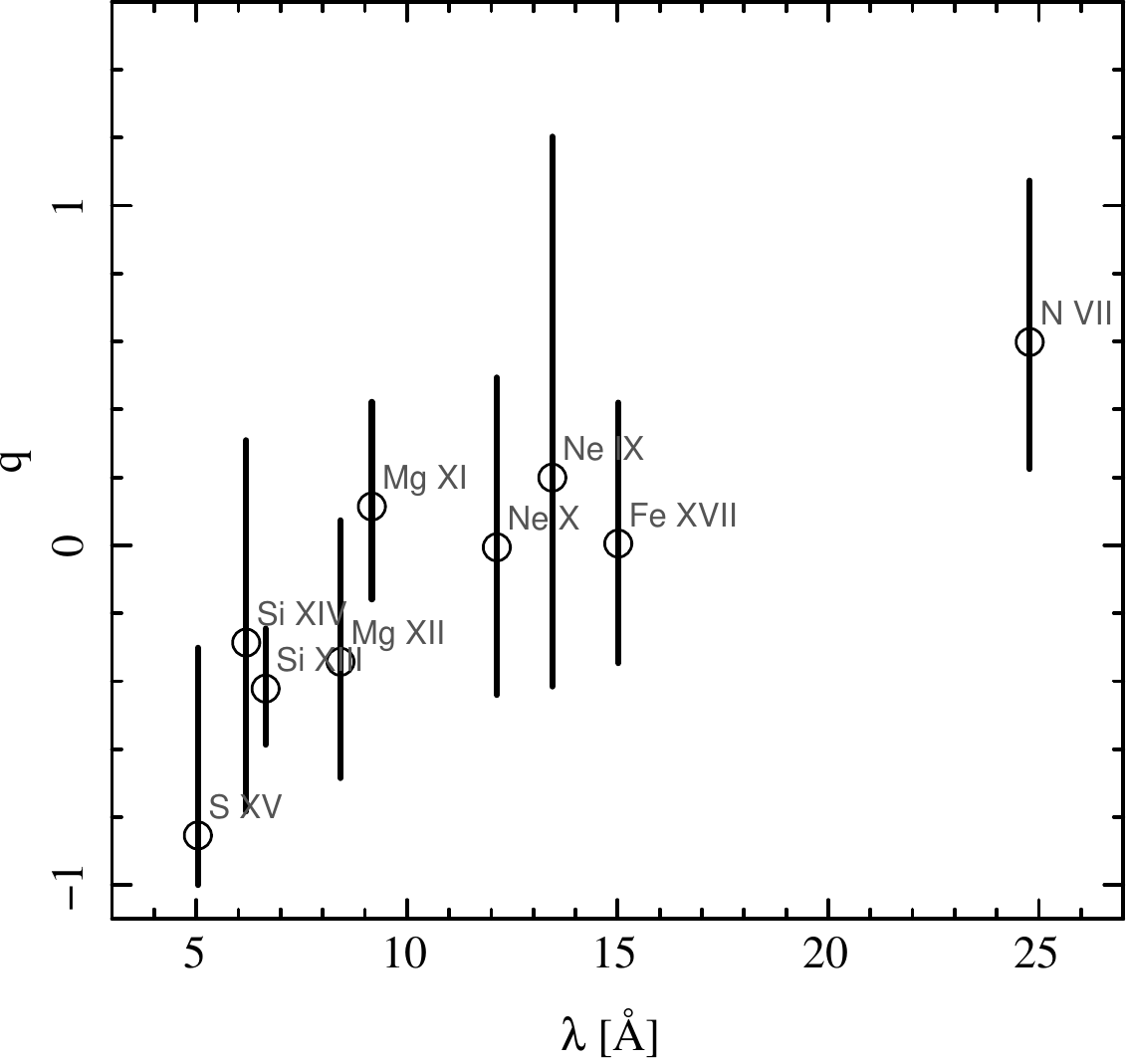}
  \caption{The line width and shape parameters against wavelength.
    On the left, there is no apparent trend of width with wavelength.
    On the right, there is a weak trend in the shape with
    wavelength. Errorbars give 90\% confidence intervals.  The points
    for \eli{N}{7} are from \xmm/RGS data.
  }
  \label{fig:qvvsw}
\end{figure}

\begin{deluxetable}{rrrl}
  \tablecolumns{3}
  \tablewidth{0.0\columnwidth}
  \tablecaption{Line Shape Parameters\label{tbl:qvinf}}
  \tablehead{
    \colhead{Line}&
    \colhead{$\lambda$}&
    \colhead{$v_\infty$}&
    \colhead{$q$}\\
    &
    \colhead{$[\mang]$}&
    \colhead{$[\kms]$}&
  }
  \startdata
  \eli{S}{15}&  $    5.04$& $1770$ $(1520, 2080)$& $ -0.86$ $( -1.00,  -0.30)$\\
  \eli{Si}{14}& $    6.18$& $1650$ $(1350, 1930)$& $ -0.29$ $( -0.78,   0.31)$\\
  \eli{Si}{13}& $    6.65$& $1980$ $(1920, 2050)$& $ -0.42$ $( -0.59,  -0.25)$\\
  \eli{Mg}{12}& $    8.42$& $1890$ $(1760, 2030)$& $ -0.34$ $( -0.68,   0.07)$\\
  \eli{Mg}{11}& $    9.17$& $2060$ $(1980, 2150)$& $  0.11$ $( -0.16,   0.42)$\\
  \eli{Ne}{10}& $   12.13$& $2050$ $(1930, 2180)$& $ -0.01$ $( -0.44,   0.49)$\\
  \eli{Ne}{9}&  $   13.45$& $1800$ $(1670, 1980)$& $  0.20$ $( -0.42,   1.20)$\\
  \eli{Fe}{17}& $   15.01$& $1900$ $(1820, 2020)$& $  0.00$ $( -0.35,   0.42)$\\
  \eli{N}{7}&   $   24.78$& $1830$ $(1700, 1960)$& $  0.60$ $(  0.22,   1.07)$\\
  \enddata
  \tablecomments{The 90\% confidence limits are given in the
    parentheses.  The values for \eli{N}{7} are from \xmm/RGS data.}  
\end{deluxetable}

\clearpage
\subsection{Light Curve Extraction}

\begin{figure}[!htb]
  \centering\leavevmode
  \includegraphics*[width=0.70\columnwidth, viewport=0 20 450 235]{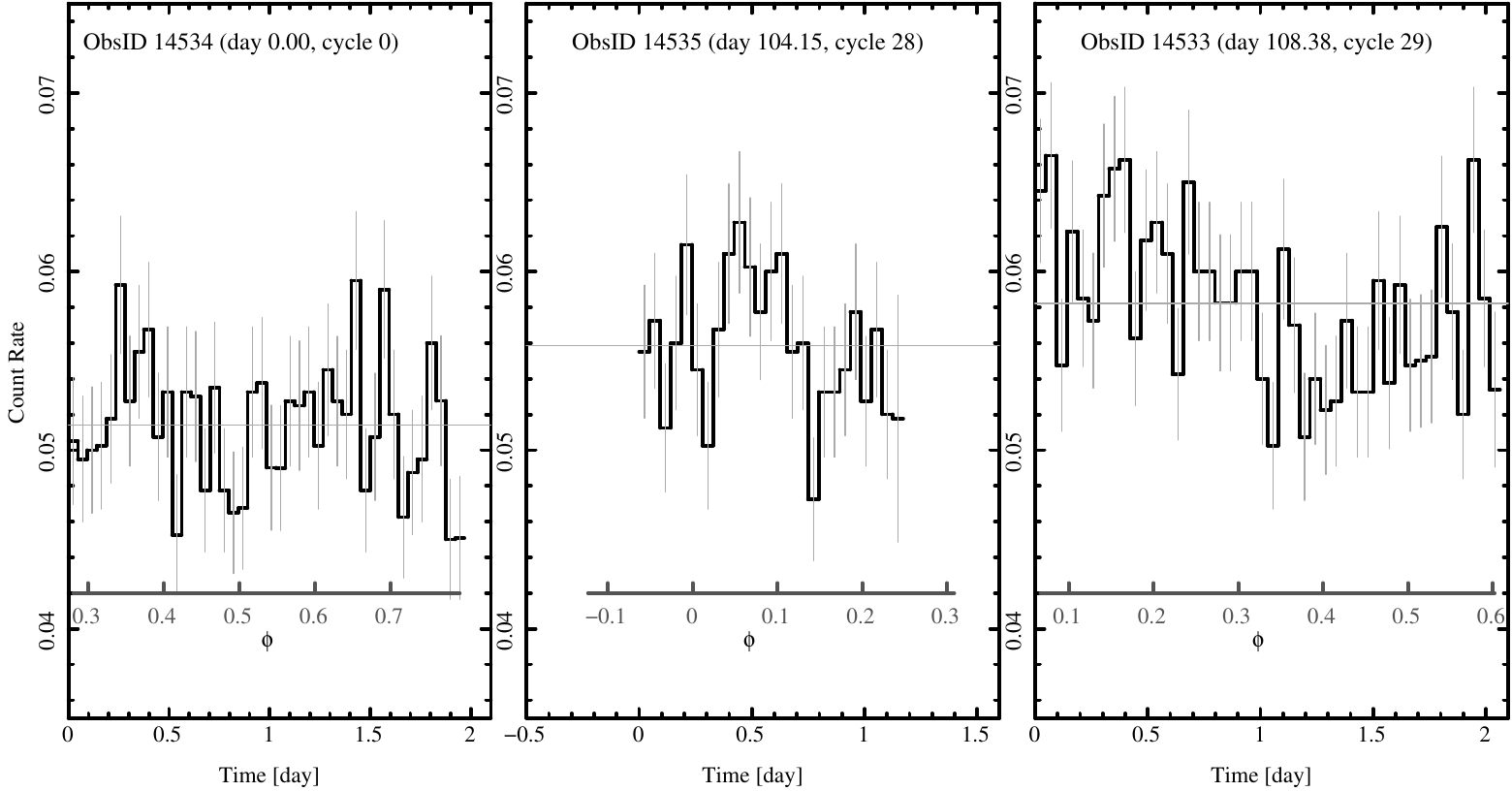}
  \includegraphics*[width=0.70\columnwidth, viewport=0  0 450 235]{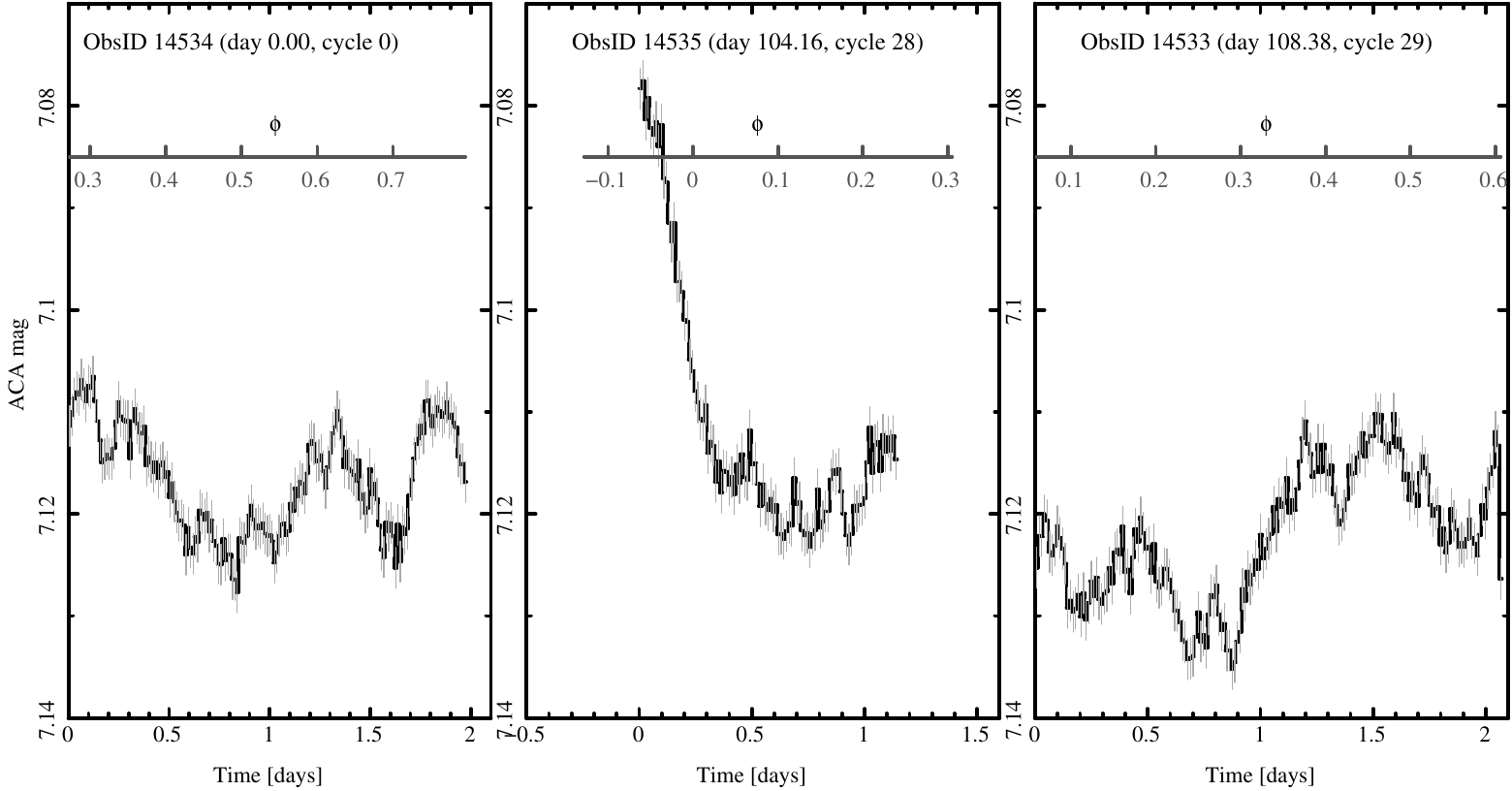}
  \caption{
    Light curves; Top: \hetg dispersed first order plus zeroth order light
    curve for each observation as labeled. The ``day'' value is the number
    of days since the $\mathrm{Time}=0$ point, based on the first
    observation, in each panel, and the cycle is the number of periods
    since the $\phi=0$ occurrence immediately preceding the start of the
    first observation.  The inset axis gives the phase according to the
    ephemeris of \citet[][(see
    Table~\ref{tbl:starparams})]{Georgiev:al:1999}. The bin size is about
    $4\ks$, and errorbars are $1\sigma$.  The horizontal gray line is the
    mean count rate for the observation.
    Bottom: the \cxo aspect camera (ACA) broad-band optical light curves in
    $1\ks$ bins taken simultaneously with the X-ray observations.    We have
    plotted $2\sigma$ uncertainties,
    estimatated to be about     $0.002\,$mag in the $1\ks$ bins, based
    on other (but flat) light   
    curves at  similar magnitude in the \chan Variable Guide Star Catalog
    \citep{Nichols:al:2010}. 
    See Table~\ref{tbl:obsparams} for observation dates.  }
  \label{fig:Lightcurves}
\end{figure}

To examine the time history of the X-ray emission, we binned
count-rate light curves from both the dispersed and zeroth order
events.  For dispersed events, we used the program, {\tt
  aglc}\footnote{ACIS Grating Light Curve, or {\tt aglc}, is available
  from \url{http://space.mit.edu/cxc/analysis/aglc/}} which handles
the dispersed photon coordinates and CCD exposure frames over multiple
CCD chips. We used first order photons from both \heg and \meg
gratings over the range $1.7$--$17.0\mang$.  For zeroth order curves,
which involve only one CCD chip, we used the \ciao program, {\tt
  dmextract}.  Count rates from the dispersed spectrum were slightly
higher than from zeroth order.  Figure~\ref{fig:Lightcurves} shows the
combined zeroth and first order rates for each of the three
observations.  Table~\ref{tbl:obsparams} lists the mean count rates
per observation and the overall means for dispersed and zeroth orders.
Hardness ratios were also computed, but since these showed no trend,
we have not included them in the figure.

Optical photometry was obtained with the \chan Aspect Camera Assembly
(ACA) simultaneously with the X-ray data.  This camera has an
approximately flat broadband response between 4000-8000\AA. Details of
this system as used for studying stellar variability can be found in
the ``\chan Variable Guide Star
Catalog''\footnote{\url{http://cxc.harvard.edu/vguide/}}
\citep[{\protect{\it VGuide}};][]{Nichols:al:2010}.  Empirical
uncertainties were estimated from the variance in flat portions of
{\it VGuide} light curves for several stars over a range of magnitues.
The optical light curve, obtained during the \chan observations, is
shown in Figure~\ref{fig:Lightcurves}.

\section{Discussion}

\subsection{Line Shapes and Ratios\label{sec:lines}}

The line profiles of \wrsix are indicative of a constant spherical
expansion.  All the lines in the \hetgs range have similar shape and
width.
%
{ 
  This implies that the radiation in all lines likely suffers from
  equally strong wind absorption, and that for each line we see only
  radiation originating from a flow with similar velocity ranges.
}
In the relatively thin-wind OB-stars, there is a strong correlation
between the line width and the wavelength.  Since the continuum
opacity is proportional to wavelength, shorter wavelength lines are
formed deeper in the wind where the expansion velocity is smaller.  In
stars like $\zeta\,$Oph or $\delta\,$Ori, there is a trend of a factor
of two or more in line widths over the \hetgs bandpass
\citep{Cassinelli:Miller:al:2001, Nichols:al:2015}.  Furthermore, the
line shape is strongly affected by material over a broad range of
velocities, and in some cases, there is evidence of temperature
stratification \citep{Herve:al:2013}.  In Figure~\ref{fig:qvvsw} we
see little or no trend in the line width with wavelength, and perhaps
a weak trend in the shape.  Hence, the evidence is strong that even
down to Si and S ($5$--$7\mang$, or to $2.5\kev$), we are only seeing
X-ray emission from large radii. {Recall that for no photoabsorption,
  $q=-1$ and a flat-topped profile would result. The weak trend in $q$
  with wavelength in Figure~\ref{fig:qvvsw} is consistent with seeing
  more emitting volume at shorter wavelengths, where the continuum
  opacity is lower.  However, it will take higher sensitivity to
  determine if this trend is real.  It will be very interesting when
  high resolution profiles can be obtained in \eli{Fe}{25} (such as
  with {\it Astro-H}) where we expect to be able to see to below
  $\sim10\,R_\ast$, to determine whether $q$ is indeed smaller, and
  also to see if the line is narrower, since at those radii we expect
  the wind velocity to be below about $\sim75\%$ of the terminal
  velocity.}

The strong UV radiation emerging from the star affects the observed
$f/i$ ratios of He-like ions, which otherwise depend solely on the
temperature of the hot plasma. By using the stellar UV field as an
input, the observed $f/i$ ratios can be used to trace the formation
region of these lines \citep{Gabriel:69, Porquet:01,
  Waldron:Cassinelli:2007}. A detailed treatment allows one to
construct an equation which predicts the $f/i$ line ratio
$\mathcal{R}$ as a function of the radiative excitation rate
$\phi_\nu$ and the electron density $n_\mathrm{e}$
\citep[cf.][Eq.\,1c]{Blumenthal:Drake:al:1972}. In the case of hot
stars, the contribution of $n_\mathrm{e}$ is almost always negligible
compared to that of $\phi_\nu$ \citep[see e.g.][]{Shenar:al:2015pp,
  Shenar:al:2015}. The simplest approach for estimating the formation
region would be to assume that the X-ray emitting gas is sharply
located at a certain radial layer, which we refer to as the formation
radius. It is more likely, however, that the X-rays originate over an
extended region. In this case, one can derive the minimal radius where
the X-ray emission of the concerned lines is originating, referred to
as the onset radius. This involves the integration of the X-ray
radiation emanating from a continuous range of radii, a method
described by \cite{Leutenegger:al:2006}, and here extended to include
the effect of K-shell absorption in the cold wind (Shenar et al. in
preparation).

\begin{figure}[!htb]
  \centering\leavevmode
  \includegraphics[width=0.45\columnwidth]{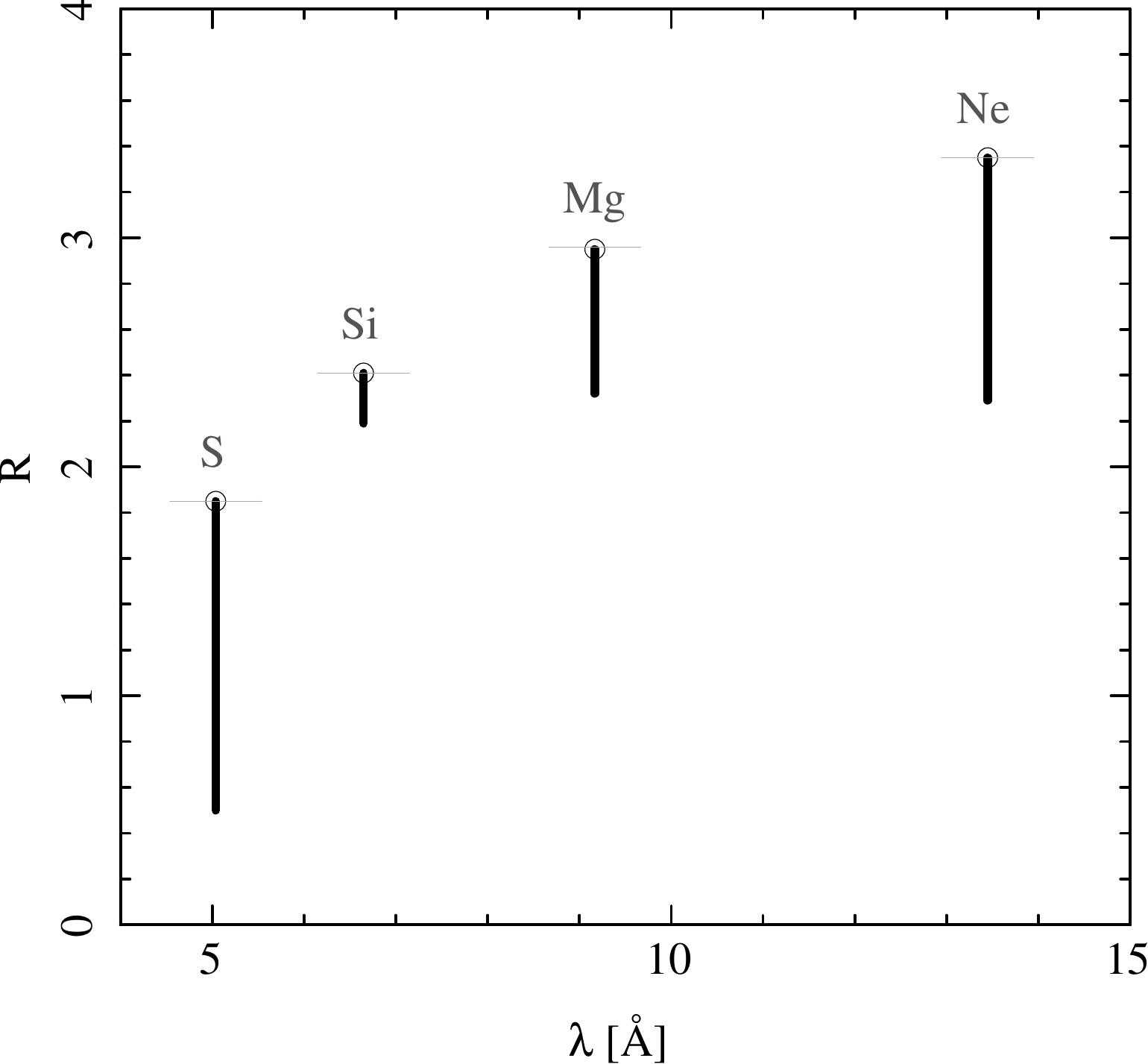}\qquad
  \includegraphics[width=0.465\columnwidth]{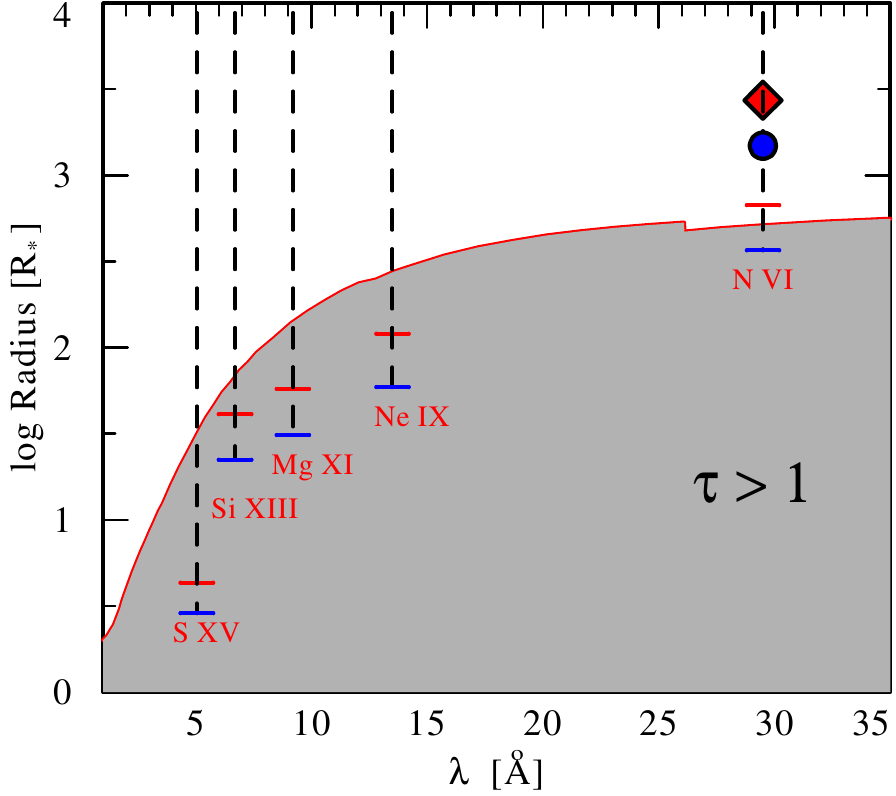}
  \caption{Left: the He-like
    $R$-ratios.  
    The circles show the best fit values, and the errorbars are 90\%
    confidence limits.  Horizontal gray segments show the maximum $R$
    for the adopted multithermal model.  Right: Using the plasma model
    $R$ fits, we show the lower limits for the He-like inferred radii of X-ray emission
    onset for distributed emission (lower, or blue bars) or
    alternatively, of formation in the localized assumption (red, or
    upper bars), versus the line wavelengths.  \eli{N}{6} is the only
    ion for which a formation radius (red diamond)      and an onset
    radius (blue circle) could be established (using    measurements
    from XMM-Newton observations).    The gray region shows 
    where the continuum opacity has $\tau_\lambda\geq1$, according to the wind
    model.
  }
  \label{fig:radfi}
\end{figure}

To determine the UV excitation rate $\phi_\nu$ and the cold wind
opacity $\kappa_\nu$, we calculated a model for \wrsix using the
non-LTE Potsdam Wolf-Rayet (PoWR) code \citep{Hamann:Grafener:2004}
with the parameters listed in Table~\ref{tbl:grparametric} as input
(also shown in Figure~\ref{fig:radfi}).  The PoWR code solves the
non-LTE radiative transfer in a spherically expanding atmosphere
simultaneously with the statistical equilibrium equations, while
accounting for energy conservation.  Complex model atoms with hundreds
of levels and thousands of transitions are taken into account, along
with millions of iron and iron-group lines which are handled using
superlevels \citep{graf2002}.  The PoWR models account for stellar
wind clumping in the standard volume filling factor micro-clumping
approach \citep[e.g.][]{hk1998}, or with an approximate correction for
wind clumps of arbitrary optical depth
\citep[macro-clumping,][]{osk2007}.  The X-ray emission and its
effects on the ionization structure of the wind is included in the
PoWR atmosphere models according to the recipe of \citet{baum1992}.
The absorption of the X-ray radiation by the relatively cool
(non-X-ray emitting) stellar wind is taken into account, as well as
its effect on the ionization stratification by the Auger process.  The
contributions of diffuse emission from the stellar wind and of limb
darkening are accounted for, and the mean intensity is accurately
calculated in the reference frame of the wind.  For the He-like lines,
the constants $R_0$ and $\phi_\mathrm{c}$ per ion, which are
calculated at a temperature at which the X-ray emission of the ion in
question peaks, are adopted from \citet{Leutenegger:al:2006}.

The right panel of Figure~\ref{fig:radfi} graphically summarizes our
results for the He-like ions \eli{S}{15}, \eli{Si}{13}, \eli{Mg}{11}
and \eli{Ne}{9}. The red bars (upper of the pair) depict the derived
90\% confidence lower limit formation radius for each He-like ion,
assuming a localized formation region. The blue bars (lower of the
pair) similarly depict the onset radii, assuming an extended formation
region. (These correspond to the lower limits of the $R$ ratios as
shown in the left-hand panel of Figure~\ref{fig:radfi}; the best fit
values yield large, unconstrained upper radii, except for \eli{N}{6}).
At radii within the gray area, the wind is optically thick to X-rays,
i.e.  $\tau_\lambda > 1$. Emission is not expected to be seen at radii
much below this surface if opacity remains high.  At first glance, it
seems that the formation regions are very distant from the
photosphere, ranging between $\approx10$ to
$\approx1000\,R_*$. However, there also seems to exist a clear
correlation between the $\tau_\lambda = 1$ surface and the formation
radii (or onset radii), especially for the \eli{S}{15} and
\eli{Si}{13} ions.  The results imply that we see the X-rays emerging
from the minimum visible radii, whereas X-rays formed below the
$\tau_\lambda = 1$ surface are absorbed.  Therefore, although the
obtained formation radii are very large, the results may actually
imply that the formation regions of X-rays originating in these ions
is much deeper---otherwise, there is no reason to expect a clear
correlation between the $\tau_\lambda = 1$ and the formation radii.
These results are also consistent with a formation radius of $\ge600$
stellar radii from the \eli{N}{6} lines seen with \xmm-\rgs
\citep{Oskinova:al:2012}.

\subsection{Line Profile Modeling Using NLTE Stellar Wind Opacity}

The same PoWR wind model as used to evaluate He-like line ratios (see
Section~\ref{sec:lines}) was used to model emission line profiles
using the detailed information about the opacity for X-rays in the
cool wind of \wrsix.  The model provides the wind ionization structure
and opacity as a function of radius, and hence the wind radial optical
depth in the X-ray band.  In this model, the ionization structure (and
consequently the mass-absorption coefficient) changes drastically with
radius. 
{
  In particular, helium recombines from double to single ionization at
  about $45\,R_\ast$.
}

It is well established that the winds of WR stars are clumped
\citep[e.g.][]{lepine1999}. Therefore, to compute the X-ray emission
line profiles, we used a $2.5\,\mathrm{D}$ stochastic wind model
\citep[see the full description in][]{Oskinova:Feldmeier:Hamann:2004, Oskinova:Feldmeier:Hamann:2006}.  For
simplicity, we assumed an idealized case of spherical clumps as well
as a constant filling factor (i.e. the case of $q=0$ of
Equation~\ref{eqn:lsf}).  The wind opacity and velocity law are
provided by PoWR models.  These models are determined by the
UV/optical spectra and are completely independent of the X-ray
spectra.  With no additional free parameters, we obtained a line
profile fully consistent with our asymptotic solution of
Equation~\ref{eqn:lsf}, and justifies the assumptions made in the
analytic profile fits, and further supports the fact that X-rays are
generated at large radii, outside the acceleration zone.

If we were to assume that X-rays were only from wind shocks in the
wind acceleration zone---as in O stars---but which in \wrsix occurs at
radii where $\tau\gg1$, then we would be seeing a very small fraction
of the total intrinsic X-ray flux produced---wind models indicate that
the optical depth of the wind for X-rays is $\tau\ge10$.
{The emergent X-rays have about $10^{-5}$ of the wind kinetic
  power ($\frac{1}{2}\dot M\,v_{\infty}^2$). If the X-rays were all
  produced in the acceleration zone (below $\sim30\,R_\ast$), then we
  would need all or more of the mechanical energy to be converted into
  X-rays to obtain what we see emerging from optical depths $>10$.}

{The quantitative models show that plasma responsible for the
  emergent X-ray line emission is distributed over the large range of
  radii between $\sim30$--$500\,R_\ast$.  The He-like lines also show
  rather directly that emission is not coming solely from deep in the
  wind.}


\subsection{Possible Reason for Extended Emission in Dense Winds}

At first glance, it might seem surprising that these winds produce
X-rays at much larger radii than seen in winds of single OB stars, as
the latter generally form their X-rays not too distant from the wind
acceleration zone where the instability is active
\citep{Krticka:Feldmeier:al:2009}.  

A much denser wind has a higher kinetic energy flux from which to
generate X-rays in shocks, and also has locally higher emission
measures at all radii.  Such a wind also has an overall higher
radiative cooling rate as a result of a higher density, so X-ray
generation should generally compete more effectively against the
adiabatic cooling of expansion.  Thus it would be possible for the
X-rays from these winds to suffer a kind of ``embarrassment of
riches,'' were it not for wind absorption.

Thus we may simply be seeing the effects of how higher density winds
shift their emission to regions where the densities are somewhat
comparable to observable X-ray generation in the winds of OB stars.
It is not obvious that a completely new mechanism is required,
although the hardness of the X-rays from \wrsix remains a puzzle,
given that the wind velocity is characteristic of OB stars as well.
All this suggests that if we are to look for a unified X-ray
generation mechanism in OB and W-R stars, it will need to be an
extended emission process that only exhibits its harder and
larger-radius components if the densities are high enough for
radiative cooling to compete successfully against expansion cooling at
such radii.

\subsection{X-ray and Optical Variability}

It has only recently become known (mainly due to the absence of
adequate data) that the intrinsic X-ray emission arising in the winds
of single hot stars is subject to slow variability of low amplitude of
typically a few percent.  Such can be seen in O-type stars
$\zeta\,$Oph, $\zeta\,Pup$, $\xi\,$Per, $\zeta\,$Ori, and
$\lambda\,$Cep \citep{Oskinova:al:2001, Naze:Oskinova:Gosset:2013,
  Massa:al:2014, Pollock:Guainazzi:2014, Rauw:al:2015}.  The 2010 \xmm
campaign on \wrsix \citep{Ignace:al:2013} revealed similar behavior of
somewhat larger amplitude over days and weeks, adding to the
historical record in which the first set of short X-ray observations
that were made with the \textit{Einstein}~Observatory between 1979 and
1981 varied in count rate by about a factor of 3 \citep{Pollock:1987}.
The long \chan observation of \wrsix is ideal for investigating X-ray
variability.  While the overall count rate changed by about 13\% over
the observing period, none of the X-ray fluctuations seen appear to be
obviously related to the 3.7650 day period of rather variable
character seen at longer wavelengths.

During the 2013 \chan campaign, judged by the overall count rates in
the dispersed and zeroth-order data, the mean intensity of \wrsix
increased by 9\% between May and August and by a further 4\% over the
3-day gap between the two final \chan observations.  These were
similar in scope to the variations seen in the \xmm campaign 3 years
earlier. With the high count rates detected with the \xmm-Newton EPIC
instruments, which exceed those of the \chan instruments by factors of
more than 30, it was possible to study in detail shorter-term
variability within an observation to show the smooth changes that
occured over several hours. We have performed a similar analysis with
the \chan data.  

The X-ray light curves are shown in Figure~\ref{fig:Lightcurves} and
display the same type of slow evolution detected with higher precision
with \xmm. At its \chan maximum at the beginning of the final \chan
observation, \wrsix was at the same brightness within the errors as
the maximum observed at the beginning of the \xmm campaign in 2010.
We did not detect any chromatic component to the variability of the
high-resolution spectra; we formed harness ratio light curves using
the first order dispersed photons between the $1.7$--$8.0\mang$
``hard'' and $8.0$--$17.0\mang$ ``soft'' bands, with null results for
variability, within the uncertainties.  We also searched for spectral
and line variability in somewhat coarser bins than the light curves
($28\ks$), with null results within the statistics.

Hence, we can only conclude that the overall flux is changing slowly,
but cannot say whether it is due to changes in X-ray temperature or
volume, or if it is due to variable absorption.

The optical light curve is also shown in Figure~\ref{fig:Lightcurves}
and presents some the most densely sampled optical photometry ever
obtained of this or any Wolf-Rayet star.  \wrsix was rapidly variable
in the optical during the X-ray observations including a rapid decline
at the beginning of the second observation and coherent structures
lasting for many hours up to as long as a day.  The X-ray variability
was much slower and of higher amplitude and bears no apparent relation
to the optical light curve.

\subsection{Evidence of Nucleosynthesis}\label{sec:nucleo}

\begin{figure}[!htb]
  \centering\leavevmode
  \includegraphics*[width=0.5\columnwidth,viewport=0 40 380 243]{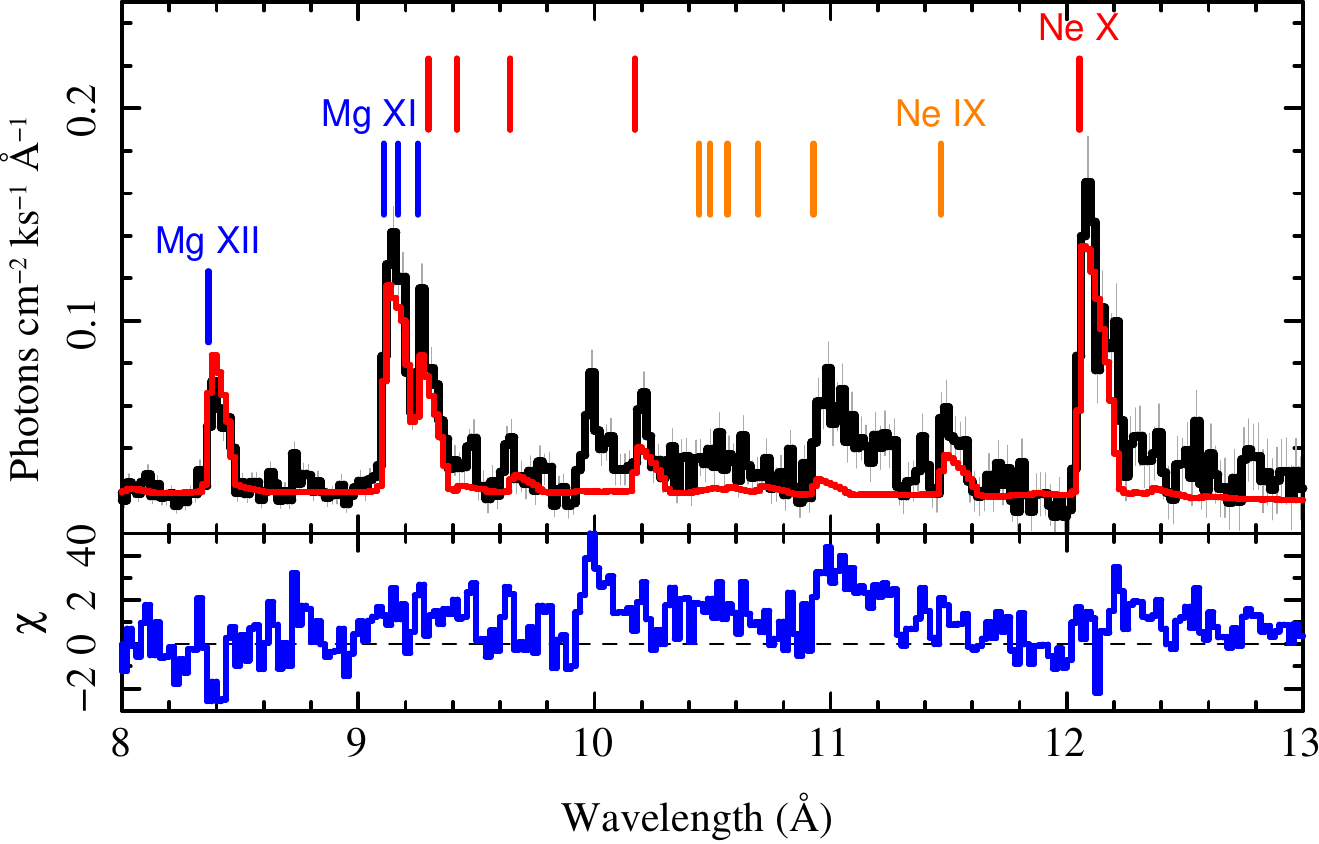}\\
  \includegraphics*[width=0.5\columnwidth,viewport=0 40 380 243]{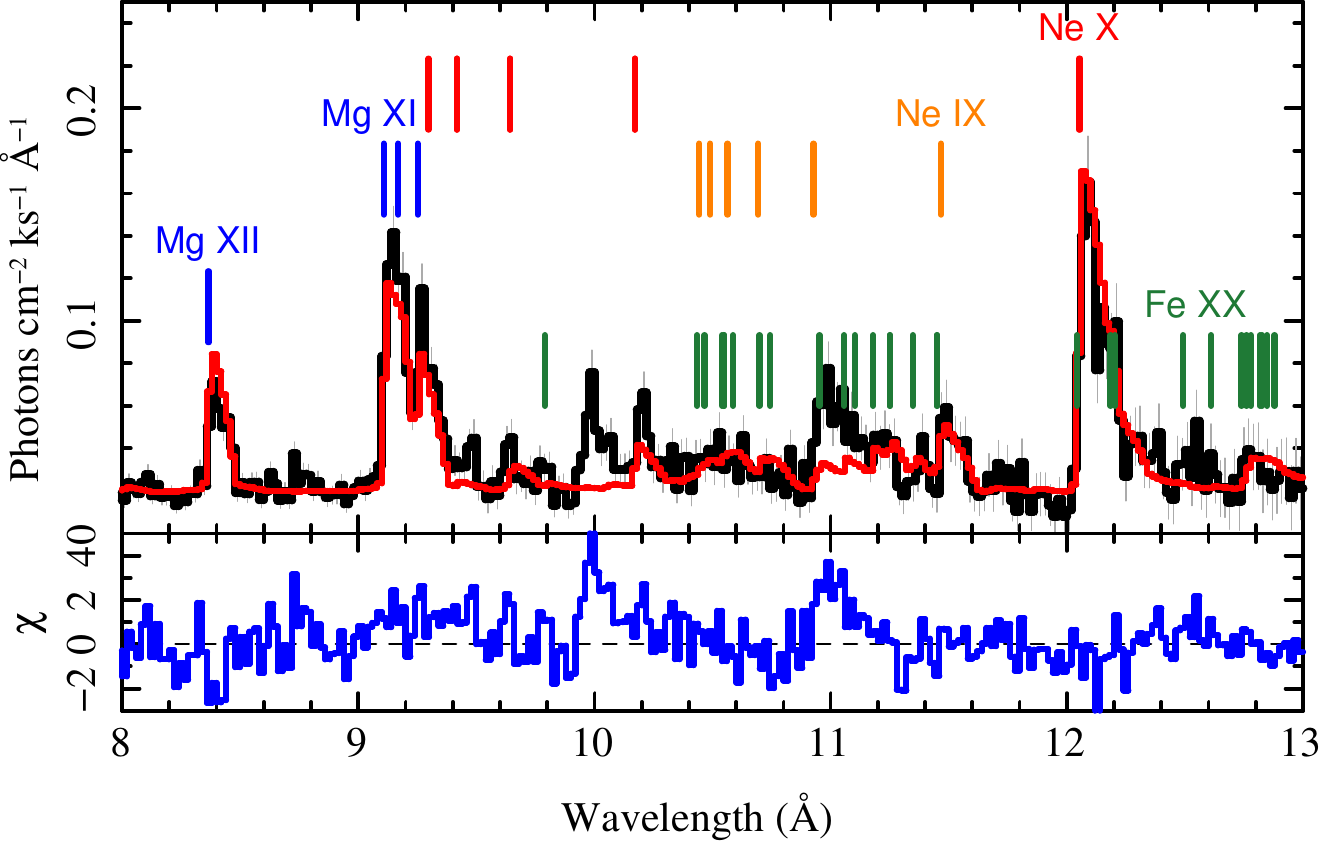}\\
  \includegraphics*[width=0.5\columnwidth,viewport=0  0 380 243]{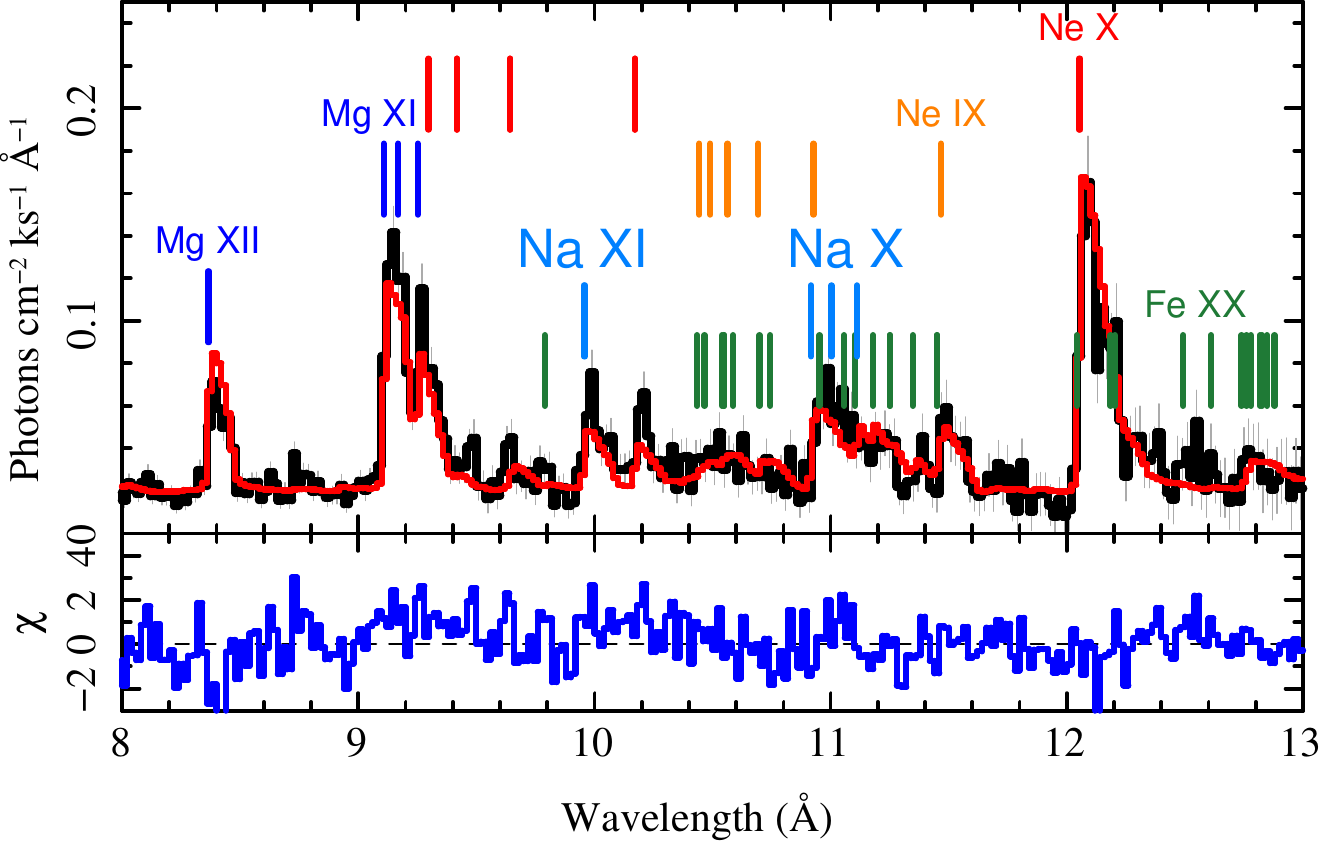}
  \caption{The region including the H- and He-like lines of
    \eli{Na}{11} ($10.02\mang$) and \eli{Na}{10} ($10.99\mang$). The
    top panel has both the Na and Fe abundances set to $0.0$.  In the
    center panel, we have restored the model's Fe abundance, and in
    the lower Na with an abundance 7 times the cosmic value. In each
    panel the black histogram is the observed spectrum, red is the
    model, and beneath in blue are the residuals.  Series of the
    brightest lines are marked at their blue edge with a different
    $y$-offset for each ion.  In the case of Fe, we marked positions
    of the 30 brightest model lines of \eli{Fe}{17}--{\sc xxiv} in the
    region (all labeled by \eli{Fe}{20}).  The reduction of the
    residuals with the inclusion of Na---at high relative
    abundance---is clear.  }
  \label{fig:sodium}
\end{figure}

Our spectral model adopted abundances from
\citet{Asplund:Grevesse:al:2009} with CNO modified according to the WN
type and \wrsix optical and UV models computed with PoWR
\citep{Hamann:Grafener:Liermann:2006}.  Abundances of a few species
with strong lines were left free in the fit to compensate for
systematic effects due to using only 4 discrete temperature components
and a simplified absorption model (see Appendix~\ref{app:specmodel}
for details).  With this model, we found that we had very large
residuals for the H- and He-like lines of sodium (\eli{Na}{11},
$10.021$, $10.027\mang$; \eli{Na}{10}, $10.990$ (resonance line),
$11.066$, $11.074$ (intercombination lines), and $11.186$ (forbidden
line)).  Re-fitting the relative abundances and model normalization in
this spectral region resulted in an overabundance of Na relative to
the abundances of \citet{Asplund:Grevesse:al:2009} by a factor of 6.9
($5.4$--$8.7$ 90\% confidence region).  We cannot explain this through
blends of Fe lines, since the many blends of \eli{Fe}{18} are formed
at about the same temperatures as other lines in the region,
\eli{Ne}{9} being the coolest having maximum emissivity near $4\mk$,
and \eli{Mg}{12} the hottest at $10\mk$.  We searched for features not
in the atomic database by looking at spectra of sharp-lined coronal
sources.  While there are features observed near $10\mang$ with no
identification in AtomDB, they are weak and not at the right
wavelength to mimic \eli{Na}{11}.  It would also be an unlikely
coincidence to have anomalies precisely at both the H- and He-like Na
wavelengths.  Hence, we conclude that the lines are due to Na and that
the abundance is enhanced.  Figure~\ref{fig:sodium} shows our best fit
to the region along with spectra without Na and without both Na and
Fe.

{ 
  Identification of Na in stellar X-ray spectra is rare, but not
  unprecedented. \citet{Sanz-Forcada:Maggio:Micela:2003} measured a
  flux for \eli{Na}{11} in \chan/\hetg spectra of AB~Dor, a young
  rapidly rotating, coronally active K-star, but did not identify the
  line.  \citet{Garcia-Alvarez:al:2005} identified \eli{Na}{11} in the
  same AB~Dor spectrum, and also in a \chan/\letgs spectrum of
  V471~Tau, a rapidly rotating K2 dwarf and white dwarf binary.
  \citet{Huenemoerder:al:2013} measured \eli{Na}{11} in the RS~CVn
  stars, $\sigma\,$Gem and HR~1099, two late-type coronally active
  binaries with among the highest fluences collected with the
  \cxo/\hetgs.  Coronally active stars have an advantage in detecting
  these weak features in that the lines are generally unresolved; the
  stars are rapidly rotating, but velocities are still below the
  instrumental resolution, and hence have relatively high contrast.
  Abundances in these cases were near solar, to within a factor of
  two.

  We do not know of any instances, however, of Na detection in other
  hot star wind spectra.  Here the broadening of the lines makes such
  a detection difficult, especially without a deep exposure or
  enhanced abundances.
}

We interpret the enhancement of Na as due to nuclear processing---the
CNO-cycle can produce Na through the Ne-Na cycle
\citep{Cavanna:al:2014}.  The amount depends upon the core
temperature, and somewhat poorly known nuclear rates
\citep{Izzard:al:2007}, but enhancement factors of 6--10 have been
predicted by models \citep{Woosley:Langer:al:1995, Chieffi:al:1998}.
The abundances of Ne, Na, Mg, and Al are all related and strongly
dependent upon the details of nuclear processing, ultimately
determined by the initial mass of the evolving star.  The soft X-ray
spectrum may be the best place to determine these abundances in
WR-stars since the emission mechanisms are relatively simple---no
radiative transfer calculations are required {in an optically
thin plasma}.  Features of Ne, Mg are strong in the \wrsix
spectrum, and Na and Al lines are present.  These certainly warrant
more careful, detailed analysis to determine their relative abundances
since they may provide unique constraints on evolutionary states.
%
%

\clearpage
\subsection{Fe~K Fluorescence}

{ 
  \citet{Oskinova:al:2012} detected Fe~K flourescence in \xmm spectra.
  In our spectra for dispersed or zeroth order, we see no
  obvious Fe~K emission next to the strong \eli{Fe}{25} emission lines
  (Figure~\ref{fig:zofek}).  The fit to the zeroth-order spectrum,
  however, is slightly improved if we introduce an additional
  component at $1.94\mang$.  The fluorescent flux we obtained
  ($3.5\times10^{-7}\pflux$, with $1\sigma$ confidence region of
  $1.7$--$5.4\times10^{-7}\pflux)$, is consistent with that measured
  by \citet{Oskinova:al:2012} in \xmm spectra, so the \hetgs spectrum
  is not inconsistent with the prior observations.

  The presence of Fe~K fluorescence requires hard photons, roughly
  $7$--$20\kev$, incident on cold Fe \citep[see][for
  example]{DrakeJJ:al:2008}.  Hence, we not only have temperatures
  high enough to produce line emission far out in the wind, there may
  also be a significant source of fairly hard continuum emission.  Our
  plasma model gives a $7.0$--$20\kev$ flux of
  $6.5\times10^{-6}\eflux$, implying a yield factor of about $0.05$.
  
  Fe~K fluorescence is interesting because it can provide constraints
  on geometry since the amount of fluorescence depends on the hard
  spectrum and on the relative locations of the cold Fe and hard
  photons.  As such, it has potential to provide another diagnostic of
  the distribution of hot and cold matter within the wind.  The
  current data, however, are not sufficient to pursue this in detail.
}

\section{Conclusions}

Our primary result, based on the resolved line profiles, is that the
X-ray emitting wind of \wrsix is undergoing constant spherical
expansion, implying a wide separation between the regions from which
the X-rays emerge ($30$--$1000\,R_\ast$) and the acceleration zone
where the line-driven instability is expected to be actively producing
shocks ($<30\,R_\ast$).  The sharp blue wings of the X-ray emission
line profiles give a good measure of a terminal velocity of $\approx
2000\kms$,
%
{
a value within the range of $1700$--$2500\kms$ determined from UV
spectra.
}
These are important conclusions, predicted by the theoretical profiles
of \citet{Ignace:2001} as very likely in WR star winds with their high
continuum opacity. These features could only be realized with a deep
exposure with the spectral resolution of \chan/\hetg.

The ratios of the He-like lines, in conjunction with the PoWR model
atmospheres (under both a simplified single radius of formation or an
onset radius with distributed emission) require that the X-rays be
produced very far out in the wind, tens to many hundreds of stellar
radii.  These locations, based on lack of photo-destruction of
forbidden lines, are consistent with the line shapes which indicate
asymptotic flow, and also with the line shape predictions derived from
the detailed opacity distribution of the PoWR models.

The X-rays from \wrsix are subject to continuous low-level variability
on different time scales. It has not yet been possible to properly
quantify this variability given its slow and seemingly stochastic
behavior, though during observations of a day or more, changes of a
few percent have been typical.  Neither the optical nor the X-ray
variations appear correlated with the known period of
$3.765\,\mathrm{d}$, nor are they correlated with each other.  It
will take the greater sensitivity of future X-ray observatories to
significantly detect and correlate such small fluctuations.

The abundance of sodium is significantly enhanced and is consistent
with probable nucleosynthesis scenarios in massive stars.  This
result is highly important for future, more sensitive high-resolution
X-ray spectroscopy, since the X-ray region may be the only place to
reliably determine the relative abundances of Ne, Na, Mg, and
Al---important elemental cycles in nucleosynthesis---in evolved stars
prior to supernovae. Obtaining such spectra should be of great
interest to new observatories such as {\it Astro-H} and {\it Athena}.

There is probably Fe~K fluorescence in the wind.  This requires hard
photons, which if thermal, implies that ``cold'' Fe is interspersed
with very hot plasma, as could be produced by strong shocks.  This hot
plasma reaches temperatures far higher---up to $50\mk$---than present
in the embedded wind shocks of ``normal'' O-stars, like $\zeta\,$Pup.

{
  The origin of such hard emission at large radii is puzzling; it may
  be a result of the way high densities at large radii allow the
  radiative cooling to compete more efficiently with other loss
  channels like expansion cooling, allowing hot gas to be seen there
  in WR-star winds but not OB-star winds.

  An optically thick wind is not going to reveal the X-rays emitted
  from embedded shocks in the acceleration region where the
  line-driven instability is active: for typical acceleration laws,
  the wind reaches 90\% of terminal velocity within 30 stellar radii,
  and for \wrsix, $\tau \gg 1$ at $30\,R_\ast$ over most of the soft
  X-ray spectrum.  Because radiative cooling is efficient in thick
  winds (and even expansion cooling acts on the scale of the radius),
  hot gas formed near the acceleration zone would cool before rising
  to radii that are optically thin to X-rays.  However, higher density
  winds extend efficient radiative cooling to larger radii, offering
  us a glimpse of mechanisms we could not otherwise detect at such
  radii. In the thinner winds of OB-stars, the density-squared
  weighting of emissivity and the rapid decline of density with radius
  means that any X-ray flux produced outside the acceleration zone is
  overwhelmed by the inner wind. In \wrsix, we can only see the outer
  wind.

  Overall, the origin of X-rays from the outer wind of \wrsix remains
  a mystery, but whether we have a new mechanism or a new view of the
  mechanisms also at work in OB-stars, observations such as this offer
  a glimpse at the spatially extended tail of the X-ray generating
  processes in hypersonic winds.  It is especially challenging to
  understand the origin of the hardest X-rays requiring the strongest
  shocks, but at least such emission can escape from deeper in the
  wind.  It suggests that some gas is accelerated to significantly
  faster speeds than the average terminal speed, and these fast
  streams can persist outside the acceleration zone, perhaps by
  passing through gaps in the clumpy slower wind.
} 

{ 
  There are a few other X-ray sources among WR-stars which are also
  thought to be single, where one can exclusively study the wind in
  X-rays, as opposed to WR-plus-O or -B binaries with strong colliding
  wind emission.  \citet{Skinner:Zhekov:al:2002a} analyzed \xmm
  low-resolution X-ray spectra of WR~110, and
  \citet{Ignace:Oskinova:Brown:2003} noted similarities among WN-stars
  WR~1, \wrsix, and WR~110 in \xmm low-resolution X-ray spectra.
  \citet{Skinner:al:2010} analyzed about a dozen more WN-stars
  observed with either \chan or \xmm. All these WN stars have thermal
  spectra, soft and hard components (though WR~1 is weak above
  $4\kev$), and are strongly absorbed. Luminosities are around
  $10^{32}$--$10^{33}\lum$.  Of note are hard components which were not
  expected and not explained by embedded wind-shock models.
  Furthermore, the lack of evidence for binary companions strongly
  excluded colliding wind shocks as an explanation.  If we place
  \wrsix on Figures 9 \& 10 in \citet{Skinner:al:2010} (using the
  \wrsix values of $\log L_\mathrm{x} = 32.9$, $\log L_\mathrm{wind} =
  37.8$, and $\log L_\mathrm{bol} = 39.2$), then it lies near WR stars
  2, 110, and 134.  In other words, \wrsix is not unusual in global
  properties compared to other putatively single WN-type WR stars.

  The X-ray spectra for all these other stars, however, are of low
  resolution. Only high resolution spectra will show whether these
  other stars share similar wind structure with \wrsix in having
  X-rays --- both hard and soft --- produced in the asymptotic uniform
  spherical outflow.  Additionally, we look forward to high resolution
  and high sensitivity of future observatories at the higher energies
  of \eli{Fe}{25} to probe the high energy processes in the deepest
  wind layers visible.
  } 
%


\acknowledgements Acknowledgements: Support for this work was provided
by the National Aeronautics and Space Administration through Chandra
Award Numbers GO3-14003A and GO5-16009A to MIT (DPH), GO3-14003D to
ETSU (RI), GO3-14003B to SAO (JN), and GO3-13003C to UI (KGG), issued by
the Chandra X-ray Observatory Center, which is operated by the
Smithsonian Astrophysical Observatory for and on behalf of the
National Aeronautics Space Administration under contract NAS8-03060.
JSN is grateful for support from the Chandra X-ray Center NASA
Contract NAS8-03060.  LO thanks the DLR grant 50 OR 1302. 

We thank Adam Foster and Randall Smith for permission and help in
running APEC for a reduced-hydrogen plasma.  We also thank Jennifer
Lauer for technical and interpretive support concerning variabiity.

{\it Facilities:} \facility{ CXO (HETG/ACIS) }

%

\clearpage
\appendix
\section{Supplemental Material}

\subsection{Plasma Model Details}\label{app:specmodel}

Figure~\ref{fig:cxoxmmfit} shows the result of a simultaneous fit of
the \wrsix spectra over all instruments.  The model was a
4-temperature AtomDB \citep{Smith:01,Foster:Smith:Brickhouse:2012}
plasma with free temperatures and normalizations, and a few free
abundances (Ne, Mg, Si, S, and Fe).  The model also included a wind
absorption component (using wind abundances), absorption from ionized
C, N, and O (C V, N IV, O IV, O V; though the oxygen and carbon
abundances are very small), and foreground interstellar absorption.
The AtomDB model was from a custom run of APEC to provide a model
without hydrogen, since this is a WN star, and otherwise using
abundances (scaled to mass fractions for a low-H atmosphere) of
\citet{Asplund:Grevesse:al:2009}, with modifications to CNO for a WN
star.  The same reduced H and CNO abundances were used for the local
(neutral and ionized wind) absorption models.

\begin{figure}[!htb]
  \centering\leavevmode
  \includegraphics*[width=0.6\columnwidth]{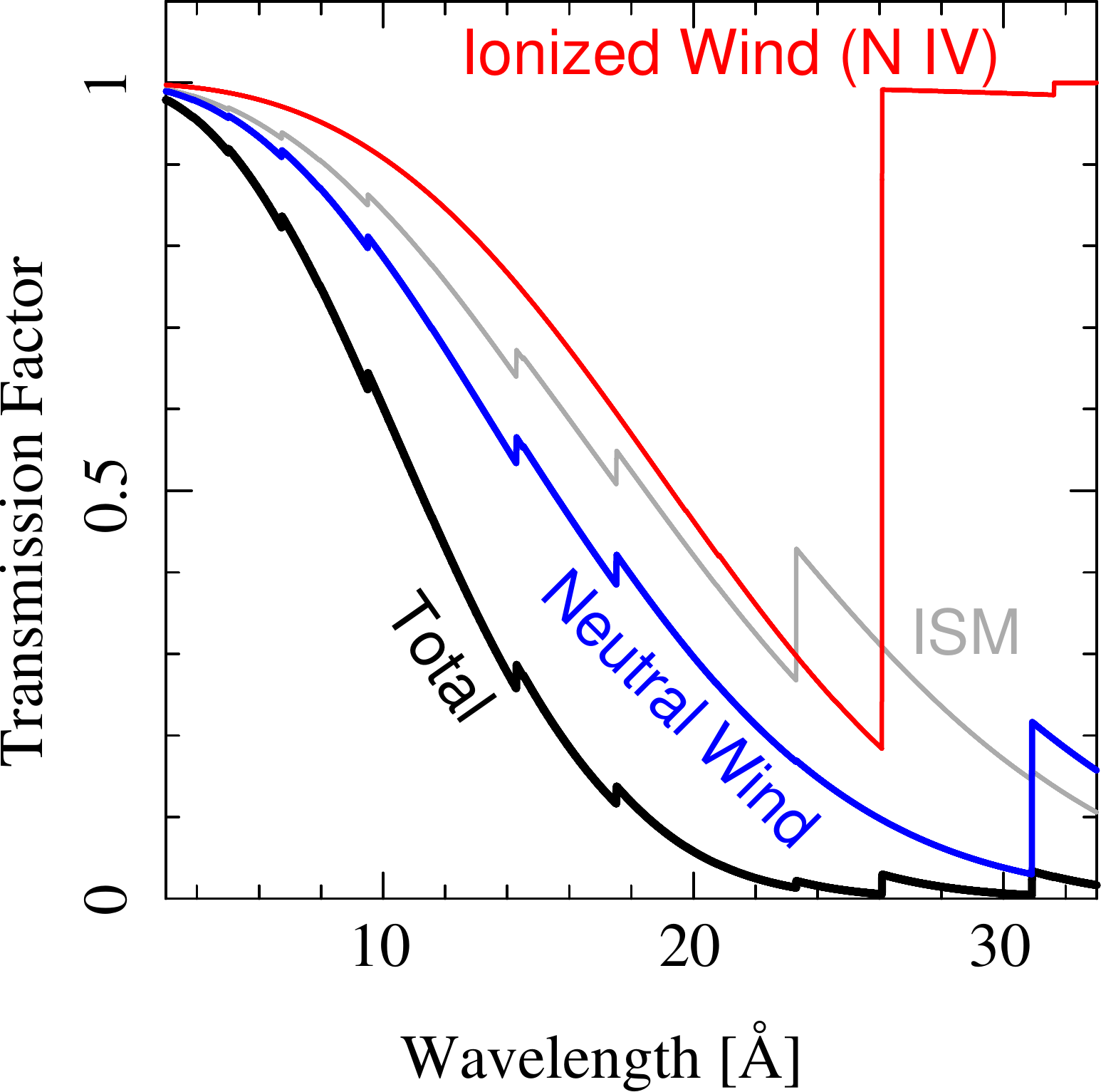}
  \caption{The components of the absorption function.  The ionized
    wind is from the {\tt XSPEC} {\tt edge} model, the neutral wind is from
    a modified {\tt XSPEC} {\tt vphabs} model, and the ISM is from the {\tt XSPEC}
    {\tt phabs} model.  The latter used standard cosmic abundances,
    while the former used the \wrsix abundances.
  }
  \label{fig:absfuns}
\end{figure}

\begin{figure}[!htb]
  \centering\leavevmode
  \includegraphics*[width=0.85\columnwidth, viewport= 30 0 570 700]{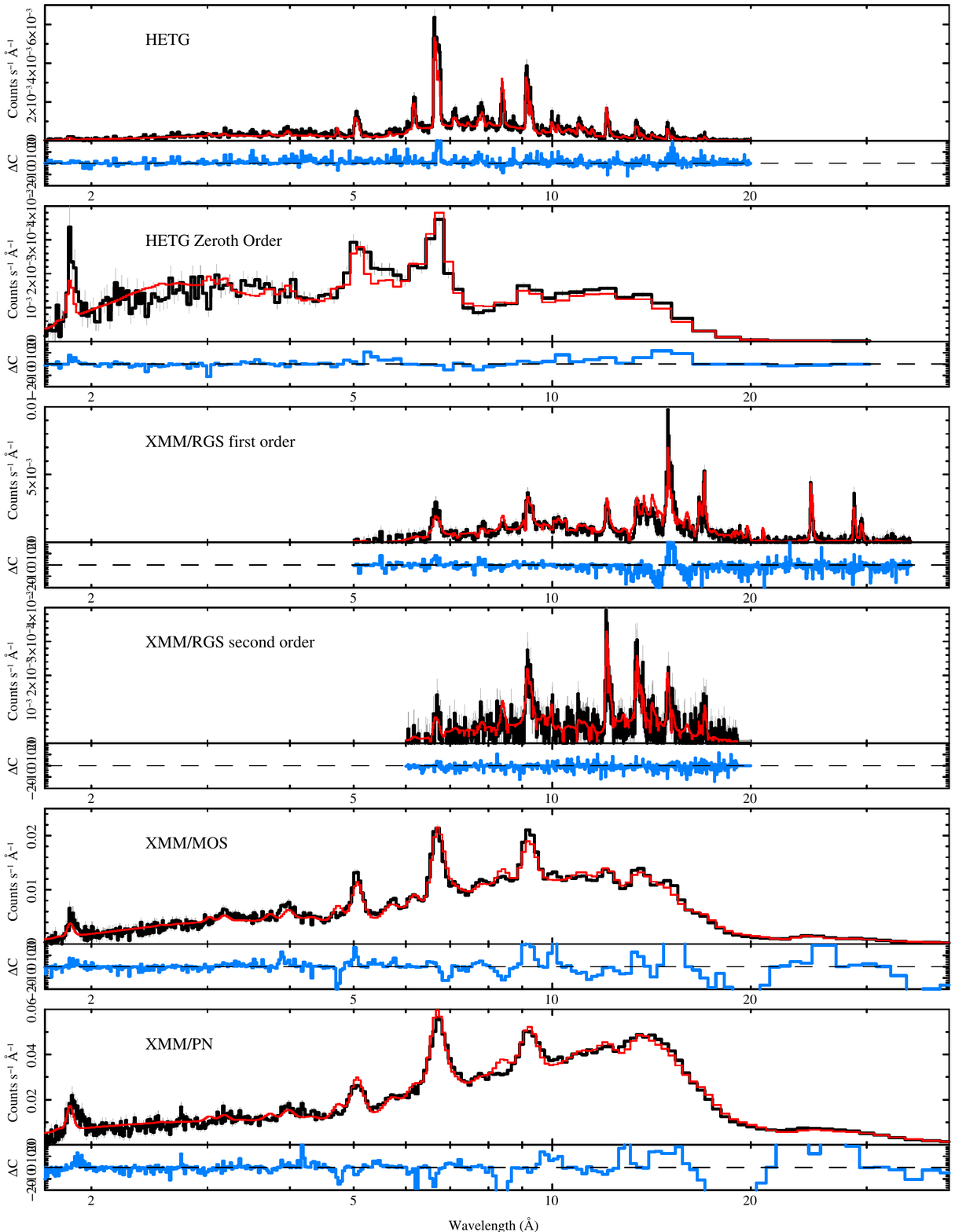}
  \caption{Plasma model compared to \chan and \xmm spectra. Black
    histograms are the observed counts, and red are the folded
    models, which are the same for each case.  Below each counts
    spectrum are residuals.
  }
  \label{fig:cxoxmmfit}
\end{figure}

The model parameters are given in
Table~\ref{tbl:modeltemps}--\ref{tbl:absmodel}.  The foreground
absorption value adopted was that used by \citet{Oskinova:al:2012},
but which they did not quote.  It is less than half that used by
\citet{Skinner:Zhekov:al:2002b}, but this is reasonable since they
only used a single absorption component for both foreground and
intrinsic components, whereas we include an intrinsic wind-absorption
component.  The absorption components are shown in
Figure~\ref{fig:absfuns}.

The fit function is effectively defined as
$${\tt Aped\_4 * windabs * vphabs\_noh * phabs}$$
where the components implemented in ISIS\footnote{See
  \url{http://space.mit.edu/cxc/isis}.} are as follows:
\begin{description}

\item[\tt Aped\_4] is a 4-component APEC model using our customized
  low-H database;

\item[\tt windabs] is a wrapper on the {\tt XSPEC} {\tt edge}
 model to provide local wind absorption by ionized CNO ions (though
 only \eli{N}{4} is significant);

\item[\tt vphabs\_noh] is a wrapper on the {\tt XSPEC} to adjust the
  model to remove hydrogen, for the wind neutral absorption function; 

\item[\tt phabs] is the unmodified {\tt XSPEC} interstellar absorption
  function.

\end{description}
In our APEC model, we also used some modifiers to implement
photoexcitation of He-like lines\footnote{\raggedright The He-like
  database and modifier as implemented in ISIS are available from
  \url{http://space.mit.edu/cxc/analysis/he\_modifier}} and to apply
the line profile model to all lines.

Abundances for Na and Al were done by fitting localized regions, as
described in Section~\ref{sec:nucleo} for Na.

%
\begin{deluxetable}{cc}
  \tabletypesize{\small}
  \tablecolumns{2}
  \tablewidth{0.0\columnwidth}
  \tablecaption{Plasma Model Components\label{tbl:modeltemps}}
  \tablehead{
    \colhead{Temperature}&\colhead{Emission Measure}\\
    \colhead{$[\mk]$}& \colhead{$[10^{54}\,\cmmthree]$}
  }
  \startdata
       1.5 &        128\\
       4.0 &         42\\
       8.0 &         24\\
      50.1 &         11\\
  \enddata  
\end{deluxetable}

\begin{deluxetable}{rrr}
  \tabletypesize{\small}
  \tablecolumns{3}
  \tablewidth{0.0\columnwidth}
  \tablecaption{Plasma Model Abundances  \label{tbl:abunds}}
  \tablehead{
    \colhead{Element} & \colhead{Fraction\tablenotemark{a}}& \colhead{Factor\tablenotemark{b}}
  }
  \startdata
  H&    -4.60&     $10^{-4}$\\
  He&    -0.01&     1.00\\
  C&    -4.01&     0.04\\
  N&    -1.82&    22.80\\
  O&    -4.01&     0.02\\
  Ne&    -3.09&     0.68\\
  Na&    -3.71&     7.00\\
  Mg&    -3.32&     0.71\\
  Al&    -4.05&     1.70\\   
  Si&    -3.09&     1.28\\
  S&    -3.07&     2.89\\
  Ar&    -4.16&     1.00\\
  K&    -5.54&     1.00\\
  Ca&    -4.22&     1.00\\
  Fe&    -3.39&     0.33\\
  Ni&    -4.65&     0.33\\
  \enddata
  \tablenotetext{a}{Decimal log of the mass fraction.}
  \tablenotetext{b}{The enhancement factor relative to the abundances
    of \citet{Asplund:Grevesse:al:2009}.}
\end{deluxetable}

\begin{deluxetable}{cc}
  \tabletypesize{\small}
  \tablecolumns{2}
  \tablewidth{0.5\columnwidth}
  \tablecaption{Absorption Model Parameters\label{tbl:absmodel}}
  \tablehead{
    \colhead{\qquad Component} & \colhead{Scale\qquad\ }
  }
  \startdata
      edge &       0.11\\
      wind &       0.20\\
       ISM &       0.17\\
  \enddata
  \tablecomments{ The ``wind'' and ``ISM'' components used modified
    {\tt XSPEC} {\tt vphabs} and {\tt phabs} models, with the hydrogen
    component scaled out.  The parameters are still in the {\tt XSPEC} units
    of equivalent hydrogen column density of $10^{22}\,\cmmtwo$.  The
    ``edge'' model used the {\tt XSPEC} {\tt edge} function; only \eli{N}{4}
    $26.085\mang$ has significant opacity, with an optical depth at
    threshold of $0.672$ per unit scale factor per abundance scale
    factor.
  }
\end{deluxetable}

\subsection{Calibration of Photon Flux \hetg Spectra}\label{app:fluxcal}

A Photon flux spectrum is an approximation to the intrinsic flux and
is obtained by dividing the counts spectrum by the model counts for a
constant flux spectrum.  It incorporates contributions to each
wavelength bin from neighboring bins as dictated by the response
matrix.  Hence, the calibration is better for near-diagonal matrices,
such as those for grating spectrometers.  Flux calibration can be
useful for visualization of a model-independent source intrinsic flux.
It does not entail any deconvolution, so the instrumental blur is
still included in the photon flux spectrum.  If lines are resolved,
however, the photon flux will approximate the {\em unconvolved} model.
We do not use photon flux spectra for quantitative analysis, but use
forward-folding methods to properly account for spectral features in
the response.  

See the ISIS manual for a detailed definition,
(\url{http://space.mit.edu/cxc/isis/manual.html}), in particular
equations 7.4--7.8.

\end{document}